\documentclass[%
 amsmath,amssymb,
 aps,
 physrev, %
]{revtex4-2}

\usepackage{graphicx}%
\usepackage{dcolumn}%
\usepackage{bm}%
\usepackage{hyperref}%

\usepackage[dvipsnames]{xcolor} %
\usepackage{adjustbox}
\usepackage{tikz}
\usetikzlibrary{calc}
\usepackage{comment}

\usepackage{etoolbox}

\AtBeginDocument{\renewcommand{\selectlanguage}[1]{}}

\begin{document}

\preprint{APS/123-QED}

\title{End-pinching and inertial-capillary reopening in viscoplastic liquid ligaments at low Ohnesorge number
}

\author{Shu Yang}
\author{Fahim Tanfeez Mahmood}
\author{C. Ricardo Constante-Amores}
 \email{crconsta@illinois.edu}
\affiliation{%
 Department of Mechanical Science and Engineering, University of Illinois Urbana-Champaign
}%

\date{\today}%

\begin{abstract}

Capillary retraction of liquid ligaments is well understood for Newtonian fluids, whereas viscoplastic effects remain comparatively unexplored. Here, we consider Herschel-Bulkley fluids, which incorporate both yield stress and shear-rate-dependent viscosity, thereby introducing a spatially varying effective viscosity that is absent  in simpler yield-stress models (e.g., Bingham models).
We focus on the low-viscosity regime, where droplet detachment in Newtonian fluids is controlled by the end-pinching mechanism.
Using fully resolved axisymmetric simulations, we show that viscoplasticity and shear-rate-dependent rheology reorganize the routes by which a retracting ligament may pinch off,  escape break-up or stay motionless due to large yield stress. 
We identify two distinct routes by which a retracting Herschel-Bulkley ligament can escape end-pinching. 
In the shear-thickening regime, increased local viscosity during neck thinning leads to larger vorticity detachment from the curved neck, which
opposes the  capillary singularity.
In the strongly shear-thinning regime, reopening is governed by curvature-induced pressure gradients.
We show that this latter mechanism persists in the Newtonian limit of vanishing viscosity, yielding a purely inertial-capillary pathway for reopening.
While previous Newtonian studies  report end-pinching  down to Ohnesorge number $Oh_K \approx  10^{-4}$,
suggesting break-up as the asymptotic low-viscosity outcome \citep{anthony_dynamics_2019}, our results demonstrate that a purely inertial-capillary reopening mechanism can arise as $Oh_K \to 0$, indicating that end-pinching is not the route in the inviscid limit.

\end{abstract}

\maketitle

\section{Introduction}

Capillary-driven ligament retraction is a canonical problem  in free-surface flows, with relevance to both  natural and   industrial processes involving liquid fragmentation and droplet formation \citep{schulkes_contraction_1996,anthony2023sharp,lohse2022fundamental,PRL_ACP}. Such dynamics underpin the formation of natural sprays and jets, and are central to technologies including drop-on-demand inkjet printing, spray coating, catalyst manufacture, and agricultural spraying \citep{shah_ink-jet_1999, eggers_physics_2008, hoath_simple_2013, anthony_dynamics_2019, planchette_breakup_2019, schena_microarrays_1998, basaran_small-scale_2002, altieri_mechanisms_2014}.

For cylindrical Newtonian ligaments, 
the retraction is dominated by a competition between capillary forces and viscosity. 
In
cases with   sufficiently low viscosity (nearly inviscid), capillary retraction results in the formation of a bulbous end at the edge which thins,  resulting in droplet detachment, this is known as  the  end-pinching mechanism \citep{schulkes_contraction_1996,notz_dynamics_2004,PRL_ACP,anthony_dynamics_2019,10.1063/1.1735095}.
\citet{hoepffner_recoil_2013} showed that viscosity could suppress the pinch-off, allowing the ligament to escape end-pinching altogether, in which increasing viscosity leads to the formation, and eventual detachment of a vortex ring generated by backflow through the neck.  
Retraction of Newtonian liquid sheets  has also been extensively studied \citep{culick_1960,taylor_1959,Savva_Bush_2009,munro2018capillary,Sanjay_Sen_Kant_Lohse_2022, Wee_2024,Ahsan}.

Despite this growing body of work, retraction dynamics involving complex rheology remain only partially understood. Existing studies have primarily focused on either interfacial effects, such as surfactant-driven Marangoni stresses, or on simple yield-stress fluids (i.e., Bingham models) in simplified geometries such as two-dimensional sheets \citep{constante-amores_dynamics_2020, kamat_surfactant-driven_2020, de_corato_retraction_2022, keshavarz_studying_2015, Sen_Datt_Segers_Wijshoff_Snoeijer_Versluis_Lohse_2021, PhysRevFluids.7.L121601,tammaro2018elasticity,tammaro2021flowering, amoroso2025numerical}. 
Surfactant-laden ligaments can produce  Marangoni stresses due to  surface tension gradients, leading to  vorticity generation  which
are sufficient to 
drive end-pinching escape, 
as demonstrated by \citet{constante-amores_dynamics_2020, kamat_surfactant-driven_2020}.
\citet{deka_retraction_2019} investigated the retraction of Bingham liquid sheets, which assume a constant post-yield viscosity, have been widely used to examine how yield stress suppresses capillary-driven motion by promoting the formation of unyielded regions.

Experimental studies of viscoplastic interfacial phenomena are commonly performed using Carbopol gels which exhibit a finite elastic modulus prior to yielding \citep{PhysRevE.73.041405, SARAMITO2009154,franca2026coalescenceprintedyieldstress}. While such elastoviscoplastic behaviour can influence interfacial dynamics, it introduces additional stress contributions associated with elastic memory {\citep{Sen_Datt_Segers_Wijshoff_Snoeijer_Versluis_Lohse_2021,PhysRevFluids.7.L121601,Zakeri_Moschopoulos_Dimakopoulos_Tsamopoulos_2025, doi:10.1021/acs.langmuir.8b00520, doi:10.1073/pnas.2105058118, Amoroso2025NumericalSO}.}
Several studies have neglected elastic effects and instead modeled the viscoplastic fluid using a Herschel-Bulkley constitutive law, such as \citet{YAO2025113550,HB_reference, MATOBA2025105481}.

Here, we adopt a Herschel-Bulkley constitutive law to model the viscoplastic fluid as a purely inelastic material with both a finite yield stress and shear-rate-dependent viscosity. Unlike Bingham fluids, whose post-yield viscosity is constant, the Herschel-Bulkley model introduces spatial variations in the effective viscosity that generate additional stress gradients and qualitatively new flow responses {\citep{10.1122/1.549926}}. As shown in section \ref{end-pinch-escape-sec}, this leads to the emergence of distinct  regimes during retraction
that are absent in the Bingham limit, where the viscosity remains uniform after yielding.
The paper is organized as follows: section~\ref{numerical-simu} outlines the numerical method and key parameter definitions; section~\ref{results_sec} explores 
the influence of yield-stress and shear-rate dependence on the dynamics and  section \ref{conclude-sec} presents the key findings and their implications, while also proposing scopes for future research.

\section{Numerical Simulations}
\label{numerical-simu}

\subsection{Governing equations}
The incompressible two-phase Navier-Stokes equations were solved employing the one-fluid formulation implemented in the open-source software Basilisk \citep{popinet_accurate_2009, popinet2013basilisk}.
Distances, times and pressures are normalized using the characteristic values: $R_0$, $(\rho R_0^3/\sigma)^{1/2}$ and $\sigma/R_0$, where $R_0$, $\rho$, and $\sigma$ represent the ligament radius, liquid density, and surface tension, respectively. Velocities were therefore scaled using the Taylor-Culick velocity  $({\sigma/\rho R_0})^{1/2}$ {\citep{taylor_1959, culick_1960}}. Applying this scaling yields the following set of dimensionless equations,
\begin{equation}
\tilde{\rho}\,
\frac{D \tilde{\mathbf{u}}}{D \tilde{t}}
=
- \tilde{\nabla} \tilde{p}
+ \tilde{\nabla} \cdot
\tilde{\boldsymbol{\tau}}
+ \tilde{\kappa}\,\tilde{\delta}_s\,\tilde{\mathbf{n}},
\qquad
\tilde{\nabla} \cdot \tilde{\mathbf{u}} = 0.
\label{eq:ns_dimless}
\end{equation}

\noindent
here, $\tilde{\mathbf{u}}$, $\tilde{p}$, $\tilde{\boldsymbol{\tau}}$, 
$\tilde{\kappa}$, $\tilde{\mathbf{n}}$, and $\tilde{\delta}_s$
correspond to the velocity field, pressure, deviatoric stress tensor, interface curvature, unit normal vector to the interface, and surface delta function, respectively. The viscosity and density functions $\tilde{\rho}$ and $\tilde{\mu}$ are defined later in this section, in equation (\ref{eq:mixture}).
In a one-fluid formulation,  the deviatoric stress depends on the local material properties. The external phase is Newtonian with constant viscosity $\mu_{\text{ext}}$ and no yield stress.
The ligament is modeled as a Herschel-Bulkley fluid, whose deviatoric stress tensor
\begin{equation}
\tilde{\boldsymbol{\tau}}
=
2\,\tilde{\mu}_{\mathrm{eff}}(\|\tilde{\boldsymbol{\mathcal{D}}}\|)\,\tilde{\boldsymbol{\mathcal{D}}},
\label{eq:HB_stress}
\end{equation}
where the dimensionless rate-of-deformation tensor,$\tilde{\boldsymbol{\mathcal{D}}},$  is defined as
\begin{equation}
\tilde{\boldsymbol{\mathcal{D}}}
=
\frac{1}{2}
\left(
\tilde{\nabla}\tilde{\mathbf{u}}
+
(\tilde{\nabla}\tilde{\mathbf{u}})^{T}
\right),
\qquad
\|\tilde{\boldsymbol{\mathcal{D}}}\|
=
\sqrt{2\,\tilde{\boldsymbol{\mathcal{D}}}:\tilde{\boldsymbol{\mathcal{D}}}}.
\end{equation}
The dimensionless apparent viscosity, $\tilde{\mu}_{\mathrm{eff}},$ for the Herschel-Bulkley fluid is given by
\begin{equation}
\tilde{\mu}_{\mathrm{eff}} =
\frac{\mathcal{J}}{2\|\tilde{\boldsymbol{\mathcal{D}}}\|+\tilde{\varepsilon}}
+
Oh_K\left(2\|\tilde{\boldsymbol{\mathcal{D}}}\|+\tilde{\varepsilon}\right)^{\,n-1},
\label{eq:HB_viscosity}
\end{equation}

\noindent
where 
$\tau_y$ is the yield stress, $K$ is the consistency index, $n$ is the flow-behavior index, and $\varepsilon$ denotes the regularization parameter. Tildes are dropped henceforth. The dimensionless groups associated with equations (\ref{eq:ns_dimless}) and (\ref{eq:HB_viscosity}) are defined as:
\begin{equation}
\mathcal{J} = \frac{\tau_y R_0}{\sigma}, \quad Oh_K = \frac{K\dot{\gamma}_c^{n-1}}{\sqrt{\rho \sigma R_0}}, 
\quad m = \frac{\mu_\mathrm{ext} }{K \dot{\gamma}_c^{n-1}},\quad \rho_r = \frac{\rho_ \mathrm{ext}}{\rho},
\label{eq:dimless_groups}
\end{equation}
where $\mathcal{J}$ denotes the plastocapillary number, measuring the competition between capillary and yield stresses and $Oh_K$ represents the Ohnesorge number, comparing inertial-capillary to inertial-viscous time scales. Moreover, $m$ is the  viscosity ratio, $\rho_r$ is the outer-to-inner density ratio, and $\dot{\gamma}_c$
is the characteristic shear-rate defined by the inertial-capillary timescale.  We note that in the limit of $n=1$ and $\tau_y=0$,  the Herschel-Bulkley model reduces to a Newtonian fluid with viscosity $K$. In all simulations, we use the ratios
$\rho_r =  10^{-3 }$ and  $m =  10^{-2}$,
which ensures that the influence of the outer medium remains minimal.

For comparison, \citet{sanjay_bursting_2021} employed a regularised Bingham formulation in which the apparent viscosity $\mu_{\mathrm{eff}}=\mu_{\min}+\mathcal{J}/(2\|\boldsymbol{\mathcal{D}}\|)$ was truncated using a prescribed maximum viscosity $Oh_{\max}$, i.e. $\mu_{\mathrm{eff}}=\min(\mu_{\min}+\mathcal{J}/(2\|\boldsymbol{\mathcal{D}}\|),Oh_{\max})$. In contrast, the present work adopts an $\varepsilon$-regularised Herschel-Bulkley model in which the strain-rate invariant is smoothed directly, yielding a fully differentiable constitutive law without viscosity clipping. Moreover, the inclusion of the power-law term $Oh_K(2\|\boldsymbol{\mathcal{D}}\|+\varepsilon)^{n-1}$ allows for  shear-thickening and shear-thinning behaviour, which are absent in the Bingham formulation. The two approaches therefore differ both in rheological class (Herschel-Bulkley versus Bingham) and in the mathematical structure of the regularisation (smooth invariant regularisation versus viscosity capping). The implementation has been  validated against the analytical solutions for steady, fully developed planar Poiseuille flow in the Newtonian, Bingham, and Herschel-Bulkley limits.  The numerical solutions collapse onto the analytical profiles, see Appendix D for more details.
Additionally, Appendix A presents a sensitivity analysis with respect to the regularization parameter, showing that the results remain unchanged over the range considered.

\subsection{Problem statement, numerical method and validation}

\begin{figure}
\centering

\begin{tikzpicture}
  \node[inner sep=0] (imgA) {%
    \begin{minipage}{\linewidth}
      \centering
      \includegraphics[width=0.8\linewidth]{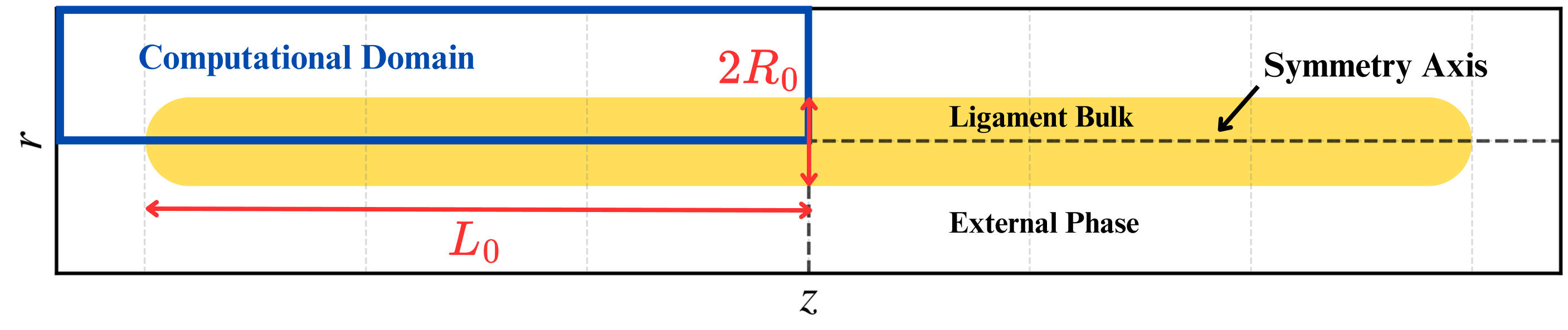}\\
      \includegraphics[width=0.7\linewidth]{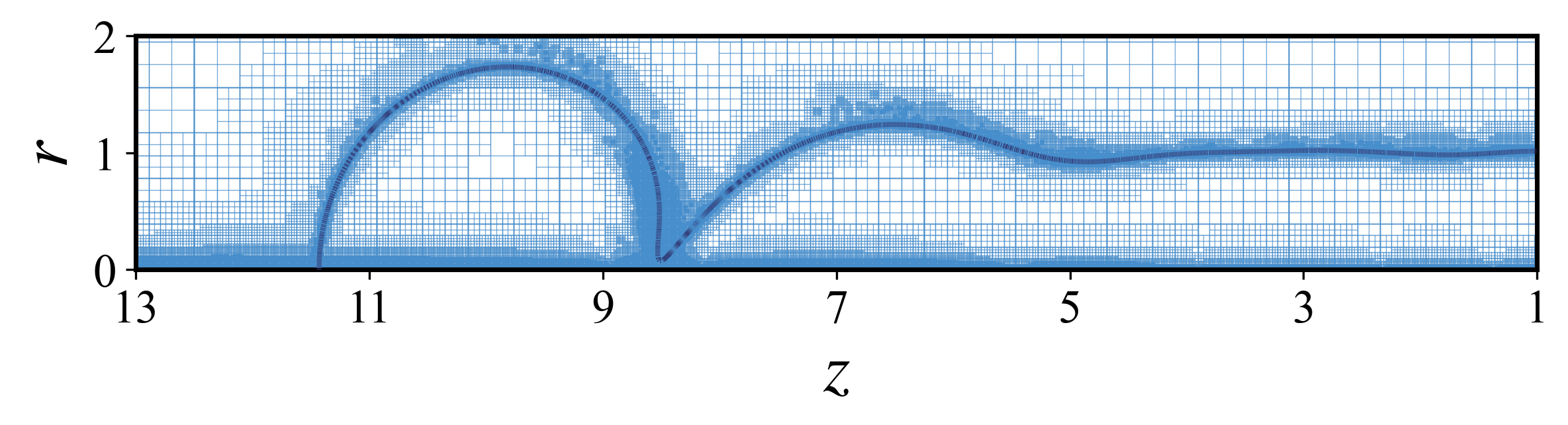}
    \end{minipage}
  };
  \node[anchor=north west,xshift=10pt,yshift=-2pt] at (imgA.north west) {\textbf{(a)}};
\end{tikzpicture}

\begin{tikzpicture}
  \node[inner sep=0] (imgB) {%
    \includegraphics[width=0.9\linewidth]{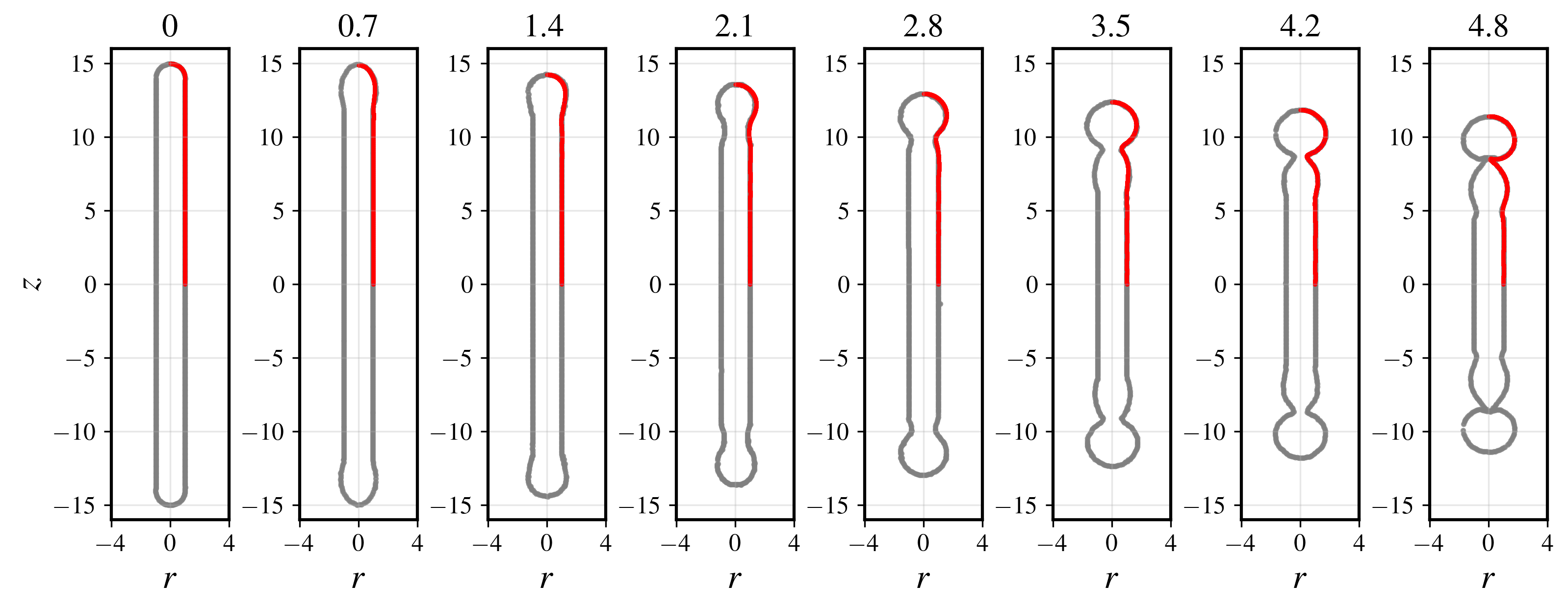}
  };
  \node[anchor=north west,xshift=-5pt,yshift=-4pt] at (imgB.north west) {\textbf{(b)}};
\end{tikzpicture}

\caption{(a) Schematic of the initial configuration of the system (top panel, not to scale) and the computational mesh used in the simulations (bottom panel). (b) Spatio-temporal evolution of a retracting ligament for $\mathcal{L}=15, Oh_K=0.001,\mathcal{J}=0$ (Newtonian). The red lines correspond to the results of the present study and the solid black lines  are collected from \citet{notz_dynamics_2004}. We note that \citet{notz_dynamics_2004} used a quarter-domain exploiting symmetry, but presented full-ligament figures.}
\label{fig:computational_domain}
\end{figure}

The system of equations~(\ref{eq:ns_dimless})--(\ref{eq:HB_stress}) is solved using the open-source solver \textsc{Basilisk~C}, which employs a finite-volume discretization on an adaptive quadtree grid, coupled with a volume-of-fluid (VOF) interface-capturing method. In the VOF framework, the interface is tracked implicitly through a tracer field $c(\tilde{\mathbf{x}},\tilde{t})$, which takes the value $1$ inside the ligament, $0$ in the outer fluid, and intermediate values $0<c<1$ across the interface.
Through this tracer field, the density and viscosity functions in equation~(\ref{eq:ns_dimless}) are defined as
\begin{equation}
\tilde{\rho} = c + (1-c)\,\rho_r, 
\qquad
\tilde{\mu} = c\,\tilde{\mu}_{\mathrm{eff}} + (1-c)\,m.
\label{eq:mixture}
\end{equation}
\noindent 
Here, $\tilde{\mu}_{\mathrm{eff}}$	corresponds to the Herschel-Bulkley apparent viscosity defined in equation (\ref{eq:HB_viscosity}), which describes the viscosity within the ligament phase ($c=1$), while the external phase is Newtonian with constant   density $\rho_\mathrm{ext}$ and viscosity $\mu_\mathrm{ext}$.
Equation (\ref{eq:mixture}) therefore represents a VOF-based interpolation of material properties, enabling a single-field formulation of the momentum equations while preserving the phase-dependent rheology.

The initial numerical setup follows closely the configuration of \citet{notz_dynamics_2004}, which used a quarter-domain exploiting symmetry. We consider a viscoplastic ligament of initial radius $R_0$ surrounded by air. The simulation is initialized with fluids at rest in the absence of gravity. The  ligament is initialized as a cylinder of aspect ratio $\mathcal{L} = L_0/R_0 = 15$ with hemispherical caps at its  ends (see figure \ref{fig:computational_domain}a). Here $L_0$ corresponds to the half-length of the ligament, similar to  \citet{notz_dynamics_2004}.  The computational domain spans $25\times 25$ (non-dimensionalized by $R_0$), with the simulated quarter-ligament initially occupying the axial range $0\leq z \leq 15$. At the outer radial boundary, we impose a zero normal pressure gradient (Neumann condition). Along the axis of symmetry, axisymmetric regularity is enforced by setting $u_r = 0$, 
$u_\theta = 0$, and $\partial p / \partial r = 0$. 
At the ligament midplane, planar symmetry is imposed through $u_z = 0$ and $\partial p/ \partial z = 0$.

We use an adaptive mesh refinement (AMR) strategy based on wavelet-estimated discretization error \citep{van_hooft_towards_2018}.
A mesh study was conducted carefully for this setup, and a maximum grid level $l=12$ was selected. 
With the current computational domain  size of $25 \times 25$,   and an adaptive refinement up to level $l=12$ yields a minimum grid spacing  $\Delta \tilde{x} = 25/2^{12} \approx 6.1 \times 10^{-3}$,  corresponding to approximately 164 grid cells across a unit length scale of the ligament ($R_0=1$).
We note that our resolution is higher than that used in viscoplastic sheet retraction of  \citet{deka_retraction_2019} (i.e., level 10), since we are aiming to capture the interfacial singularity. The velocity and pressure fields are advanced in time by Basilisk using a classical time-splitting projection method \citep{popinet_accurate_2009}. Appendix B presents a grid-dependence study for a representative reopening case, demonstrating mesh-independent results. We also note that, for the extreme low-viscosity limit $(Oh_K \rightarrow 0)$ cases, simulations were  performed with a grid level of $l=13$ ($\approx 328$ grid cells across the radius of the ligament).

Figure~\ref{fig:computational_domain}b compares our numerical predictions with the results of \citet{notz_dynamics_2004} and demonstrates excellent agreement in the spatio-temporal evolution of the interface shape.
For this validation case, we recover the Newtonian limit by setting $\mathcal{J}=0$, $n=1$, and $\varepsilon=0$,  so that the effective viscosity reduces to a constant. Only one quarter of the thread was simulated by exploiting axial and planar symmetries, consistent with the setup of \citet{notz_dynamics_2004}. The superimposed red profiles correspond to this quarter-domain representation. The predicted pinch-off time is 
$t_b=4.857$, in close agreement with the reported value $t_b=4.797$.  The quantitative agreement therefore confirms that the Herschel-Bulkley solver correctly reproduces the  Newtonian limit.

In the current study, we restrict our attention to the primary retraction dynamics and the neck evolution, and we do not consider subsequent secondary break-up processes. When break-up does not occur, we continue the simulations only until the viscoplastic droplet fully forms, limiting the total integration time to less than 10 nondimensional units.

\section{Results}
\label{results_sec}

\begin{figure}
\centering
\includegraphics[width=\linewidth]{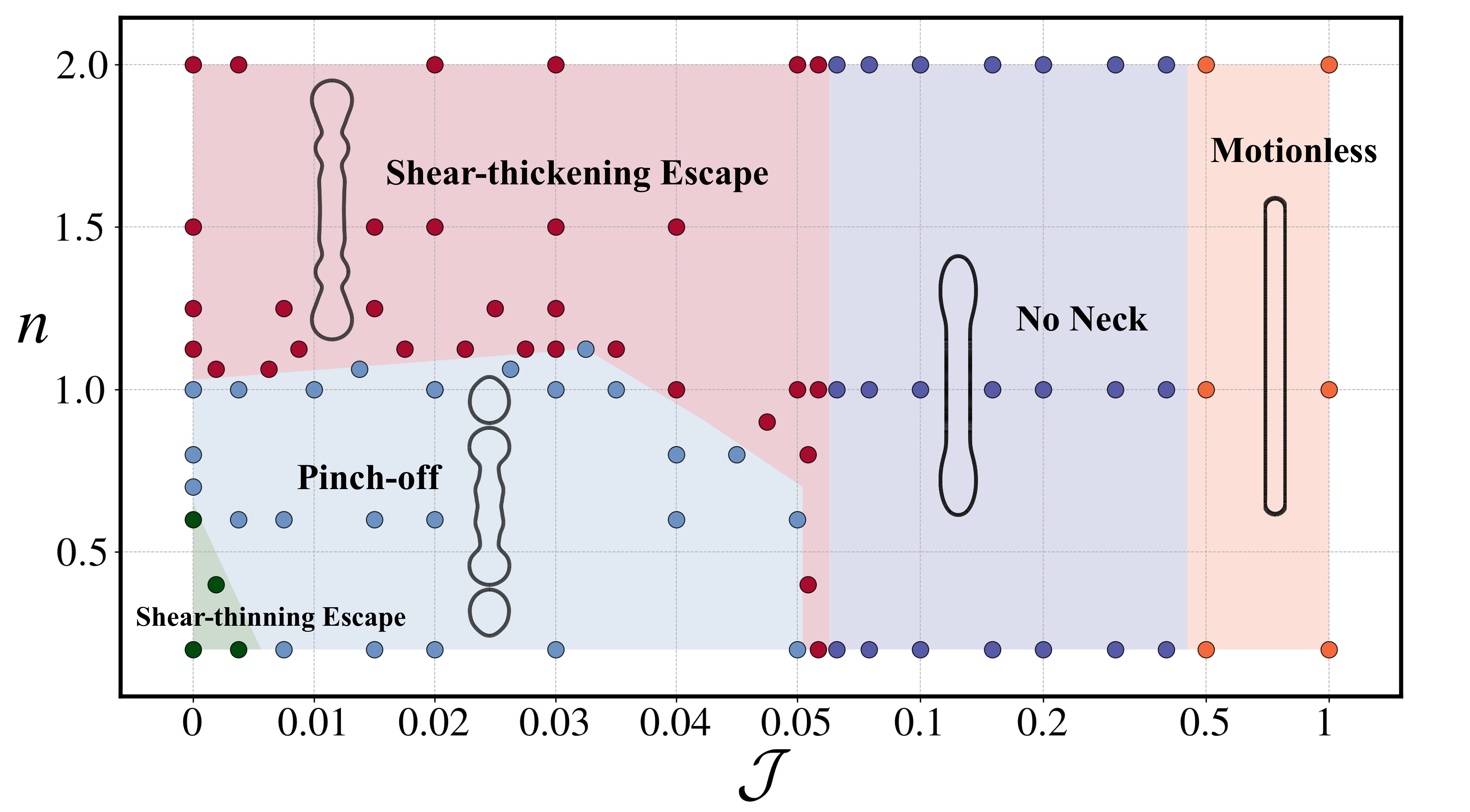}
\caption{
Regime map in terms of the plastocapillary number $\mathcal{J}$ and the flow index parameter $n$ for $Oh_K=10^{-3}$ showing the transitions between the different categories identified in the current study. The insets show the interfacial topology for representative cases from each  regime.
}
\label{fig:regime_map}
\end{figure}

The interplay between capillarity, rheology, and inertia gives rise to
 distinct outcomes such as  pinch-off, end-pinching escape, retraction without the formation of a neck,  and arrested motion. 
Their organization in the $(n,\mathcal{J})$ plane is governed by the evolving local Ohnesorge number   $Oh_{\text{loc}}(t) = \tilde{\mu}_{\mathrm{eff}} (t) / \sqrt{\rho \sigma r_{\min}(t)}$, where the dimensionless apparent viscosity $\tilde{\mu}_{\mathrm{eff}}$ is evaluated at the ligament centerline  at the axial location of the minimum neck radius.
This dimensionless quantity characterizes the instantaneous competition between viscous, capillary, and inertial stresses, 
and thus provides a direct measure of the local flow regime during thinning.

\subsection{Phenomenological regime map in  $n$ and $\mathcal{J}$ space \label{regime_map_section}}

We begin the discussion of the results by presenting a phenomenological regime map in the $(n, \mathcal{J})$ parameter space at $Oh_K = 10^{-3}$ and $\mathcal{L} = 15$. Figure~\ref{fig:regime_map} illustrates the distinct  outcomes and their corresponding boundaries. Approximately 100 simulations were performed to delineate these boundaries accurately. Each point  on the map was classified by monitoring the neck radius over time. 
Four primary regimes are identified, each characterized by qualitatively different interfacial behavior.
For shear-thinning and weakly shear-thickening fluids at small to moderate values of $\mathcal{J}$ (i.e., $\mathcal{J} \lesssim 0.045$), the break-up follows the well-known end-pinching mechanism, where a droplet detaches from the tip of the retracting ligament.
The regime map reveals that end-pinching escape persists across two rheological limits depending on whether the fluid exhibits  shear-thickening and shear-thinning behaviors.
In the intermediate range $0.06 \lesssim \mathcal{J} \lesssim 0.45$, we identify a no-neck regime, in which no localized neck forms during retraction. Instead, extensive unyielded regions
suppress axial localization, and the filament retracts smoothly into a single droplet without pinch-off. Finally, for $\mathcal{J} \gtrsim 0.45$, a motionless regime emerges in which retraction is entirely suppressed by yield stress, e.g., yield stresses dominate over the viscous and inertial contributions. These last two regimes are largely insensitive to variations in the flow-behavior index $n$ because the dominant balance is controlled by the yield stress rather than the shear-dependent viscosity.

\subsection{End-pinching regime}
\label{pinch-off-regime-sec}

\begin{figure}
\centering
\begin{tikzpicture}
  \node[inner sep=0] (A) {\includegraphics[width=\linewidth]{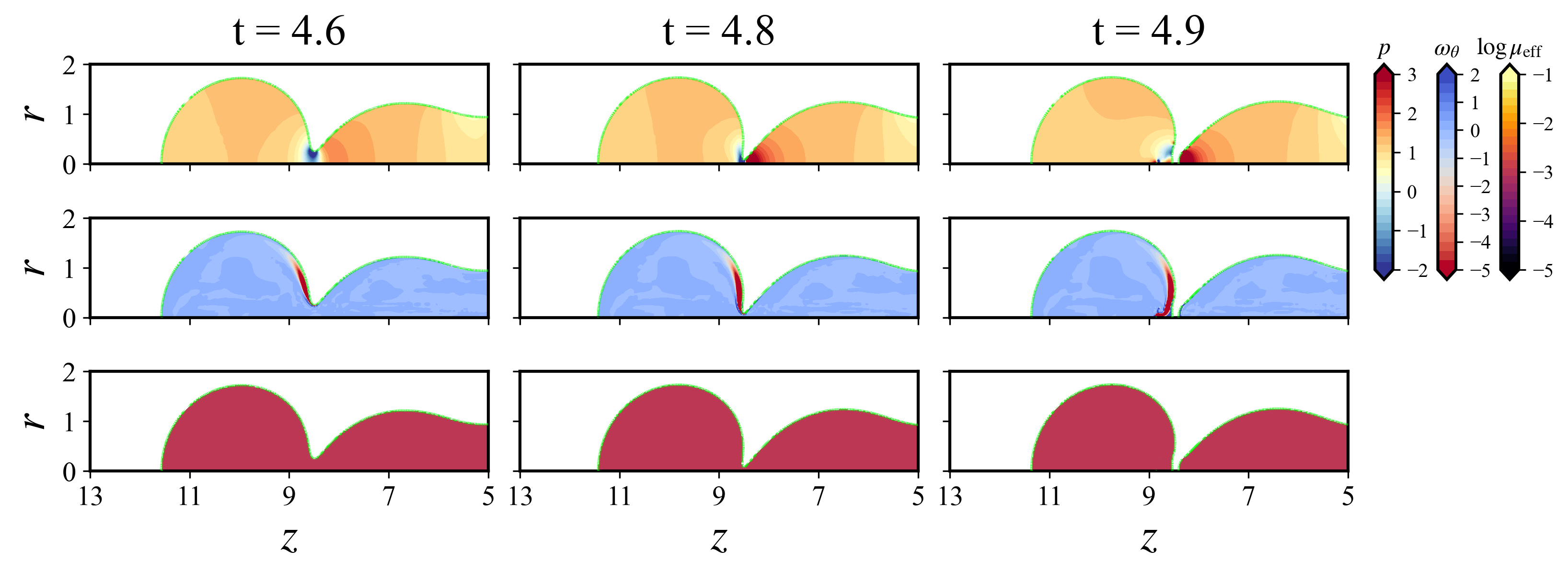}};
  \node[anchor=north west,xshift=-4pt,yshift=-2pt] at (A.north west) {\textbf{(a)}};
\end{tikzpicture}
\begin{tikzpicture}
  \node[inner sep=0] (B) {\includegraphics[width=\linewidth]{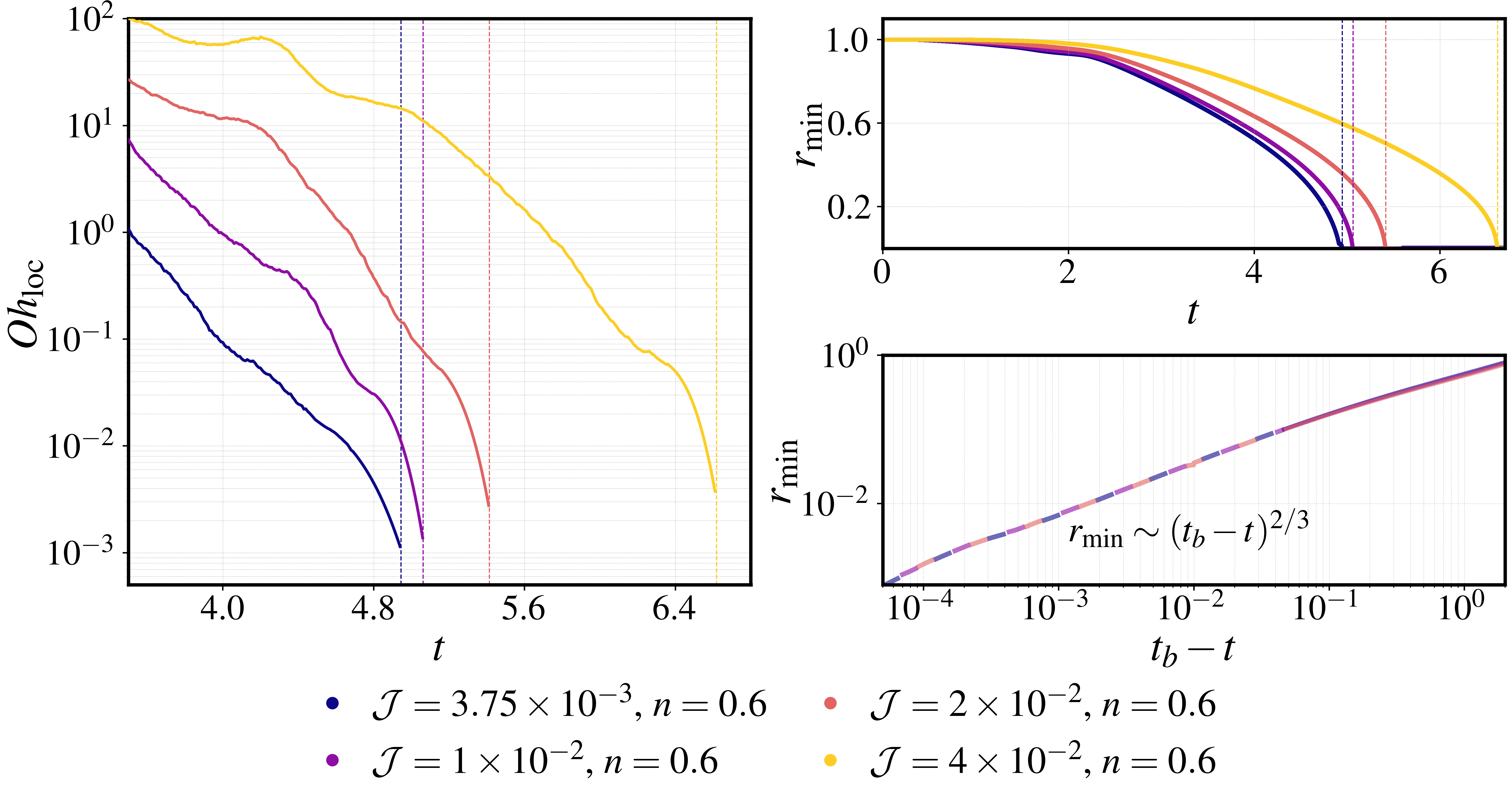}};
  \node[anchor=north west,xshift=-2pt,yshift=2pt] at (B.north west) {\textbf{(b)}};
  \node[anchor=north,xshift=10pt,yshift=2pt] at (B.north) {\textbf{(c)}};

  \node[anchor=north,xshift=10pt,yshift=-2.6cm] at (B.north) {\textbf{(d)}};
\end{tikzpicture}

\caption{
End-pinching regime.
(a) Temporal sequence of the pressure field $p$ (top), azimuthal vorticity $\omega_\theta$ (middle), and effective viscosity $\log_{10}(\mu_{\text{eff}})$ (bottom) during the final stages of break-up, shown at $t = 4.6$, $4.8$, and $4.9$ with $Oh_K=10^{-3}$, $\mathcal{J}=0$ and $n=1$
(b) Temporal evolution of  $Oh_{\text{loc}}$ at the neck for several $\mathcal{J}$ at fixed $n = 0.6$. The  break-up time is indicated by vertical dashed lines.
(c) Temporal evolution of the minimum radius.
(d)  Evolution of the minimum radius as a function of the remaining time to break-up, $t_b - t$,  shown on log-log axes.
}
\label{fig:base_pinchoff}
\end{figure}

Figure~\ref{fig:base_pinchoff}a illustrates the classical end-pinching mechanism for a Newtonian ligament at $Oh_K = 10^{-3}$ ($\mathcal{J}=0$, $\varepsilon=0$, $n=1$). A capillary pressure gradient drives fluid from the ligament bulk toward the tip, where axial deceleration leads to local mass accumulation and the formation of a bulbous end. The associated axial flux induces the formation of a neck connecting the bulb to the filament, which thins continuously in time.
The flow structure reveals that capillary pressure becomes strongly localized within the neck region as thinning proceeds, while vorticity remains confined to a thin azimuthal layer near the interface (figure~\ref{fig:base_pinchoff}a).  This Newtonian reference case establishes a reference inertial-capillary regime against which the non-Newtonian dynamics can be assessed.

The local Ohnesorge number for varying $\mathcal{J}$ at fixed $n=0.6$, $Oh_{\text{loc}}$ (figure~\ref{fig:base_pinchoff}b), decreases during thinning, indicating that viscous stresses become progressively weaker relative to inertial--capillary forces. As a result, the dynamics approach the inertial--capillary similarity regime. 
Figures \ref{fig:base_pinchoff}b-c
show that increasing  $\mathcal{J}$ merely delays the onset of this regime. 
Figure~\ref{fig:base_pinchoff}d shows that the minimum neck radius follows 
$r_{\min}\!\sim\!(t_b - t)^{2/3}$, 
consistent with classical inertial--capillary theory \citep{day_self-similar_1998}. This demonstrates that, after yielding, the flow recovers the classical self-similar regime.

\subsection{Escape from end-pinching}
\label{end-pinch-escape-sec}

In this section, we identify two distinct mechanisms governing escape from end-pinching, depending on whether the fluid exhibits shear-thickening  or  shear-thinning behaviour.

\subsubsection{Shear-thickening fluids}

\begin{figure}
\centering

\begin{tikzpicture}
  \node[inner sep=0] (A) {\includegraphics[width=\linewidth]{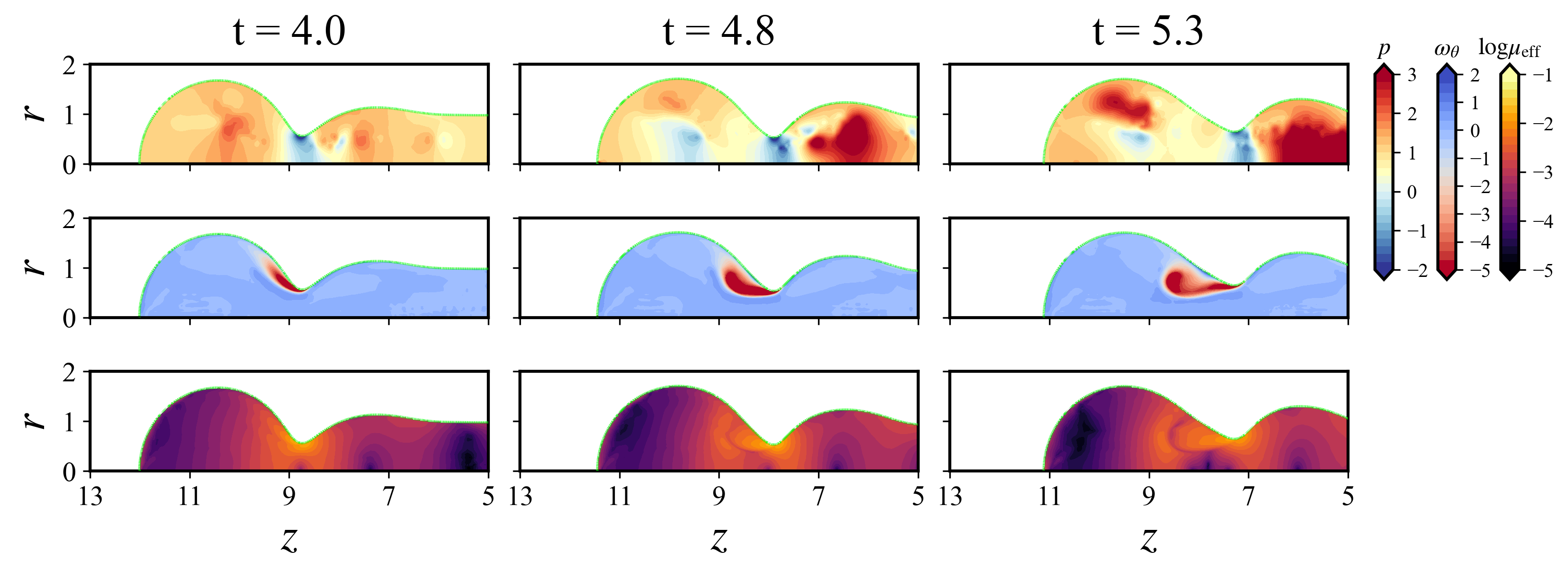}};
  \node[anchor=north west,xshift=-4pt,yshift=-2pt] at (A.north west) {\textbf{(a)}};
\end{tikzpicture}
\begin{tikzpicture}
  \node[inner sep=0] (B) {\includegraphics[width=\linewidth]{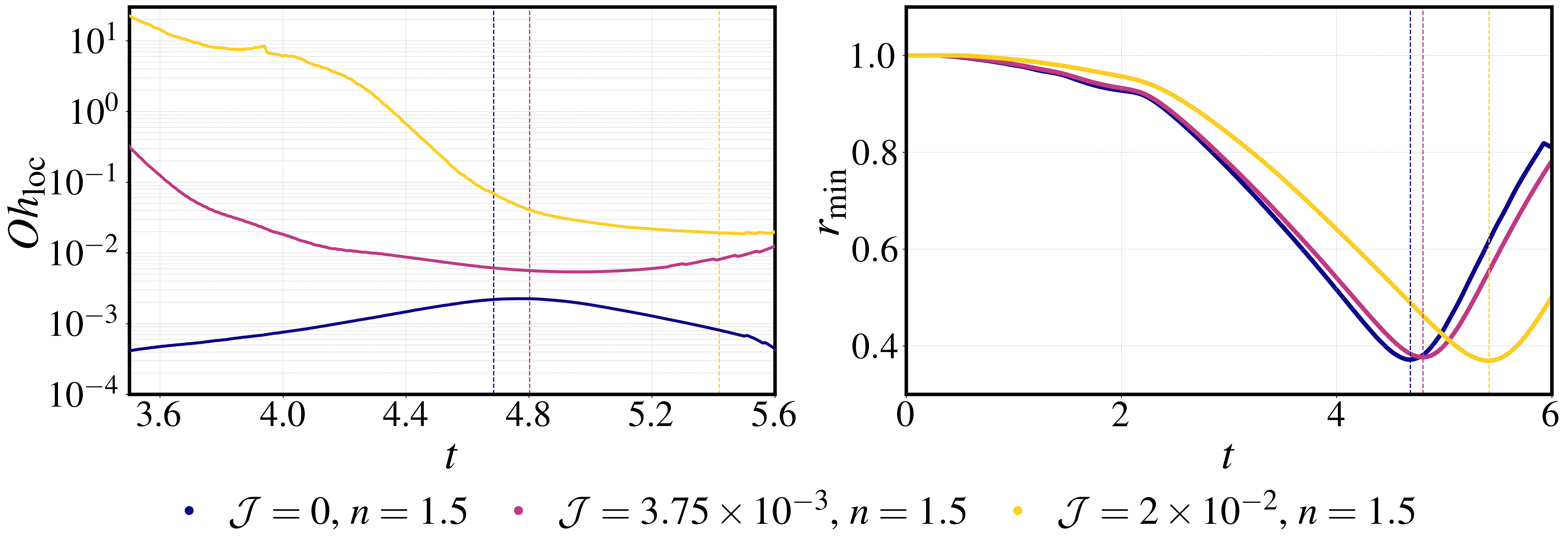}};
  \node[anchor=north west,xshift=-2pt,yshift=2pt] at (B.north west) {\textbf{(b)}};
  \node[anchor=north,xshift=10pt,yshift=2pt] at (B.north) {\textbf{(c)}};
\end{tikzpicture}

\caption{Shear-thickening reopening regime.
(a) 
Snapshots of pressure $p$, azimuthal vorticity $\omega_\theta$, and $\log_{10}(\mu_{\mathrm{eff}})$ for the case $Oh_K = 10^{-3}$, $\mathcal{J}=0$, and $n=1.5$, with each row showing one of the three fields at times  $t = 4.0$, $4.8$, and $5.3$ from left to right.
(b) Temporal evolution of the local Ohnesorge number $Oh_{\text{loc}}$ at the neck. 
(c) Minimum radius $r_{\min}$ versus time for several $\mathcal{J}$ at fixed $n = 1.5$. The position of the minimum ligament radius, $r_{min}$, is indicated by vertical dashed lines.}
\label{pinchoff_shear_thickening}
\end{figure}

For shear-thickening fluids, we observe reopening of the neck without the formation of a capillary singularity for $\mathcal{J} \lesssim 0.06$. This behavior arises from the strong increase of the apparent viscosity with the local strain rate, $\mu_{\mathrm{eff}} \sim \dot{\gamma}^{\,n-1}$ with $n>1$.

The flow dynamics underlying this process are illustrated in figure~\ref{pinchoff_shear_thickening} through the evolution of the azimuthal vorticity, pressure, and apparent viscosity fields for $Oh_K = 10^{-3}$, $\mathcal{J}=0$, and $n=1.5$. At early times (see $t=4.0$ in figure \ref{pinchoff_shear_thickening}a), the flow around the highly curved neck leads to the generation of weak vorticity in its vicinity. As time moves forward, a large concentration of vorticity develops on the bulb side of the neck, where high curvature and rapid thinning generate strong fluid motion both normal and tangential to the interface (see $t=4.8$).
Vorticity generation is accompanied by the formation of a localized interfacial shear layer with strong tangential flow, leading to flow separation. The resulting backflow within this layer advects fluid   towards the neck, driving its reopening (e.g., tis mechanism is similar to \citet{hoepffner_recoil_2013} for Newtonian viscous fluids and \citet{kamat_surfactant-driven_2020,constante-amores_dynamics_2020} for surfactant-laden threads). After escape from end-pinching, at $t = 5.3$, viscous diffusion spreads vorticity into the bulk, thickening the shear layer and leading to detachment of the vortex ring. 
Figure \ref{pinchoff_shear_thickening}b shows that $Oh_{\text{loc}}$ increases during thinning for $\mathcal{J}=0$ and $n=1.5$, reflecting the growth of effective viscosity with strain rate and indicating that reopening is mediated by viscous shear-layer dynamics.

The role of the plastocapillary number $\mathcal{J}$ is primarily to delay the onset of this dynamics. Across all cases, the neck thins to a comparable minimum radius, $r_{\min} \approx 0.4$, indicating that reopening is governed by shear-thickening rheology rather than yield stress (see figure \ref{pinchoff_shear_thickening}c). Increasing $\mathcal{J}$ suppresses early deformation by keeping low shear rates, and hence low $\mu_{\mathrm{eff}}$. Once the capillary pressure exceeds the yield threshold, the strain rate increases abruptly, leading to a rapid growth of $\mu_{\mathrm{eff}}$ and a corresponding amplification of viscous stresses. This delayed but stronger viscous response results in a larger effective resistance during late-stage thinning, without significantly altering the geometric evolution of the neck.

\subsubsection{Shear-thinning fluids}

\begin{figure}
\centering

\begin{tikzpicture}
  \node[inner sep=0] (A) {\includegraphics[width=\linewidth]{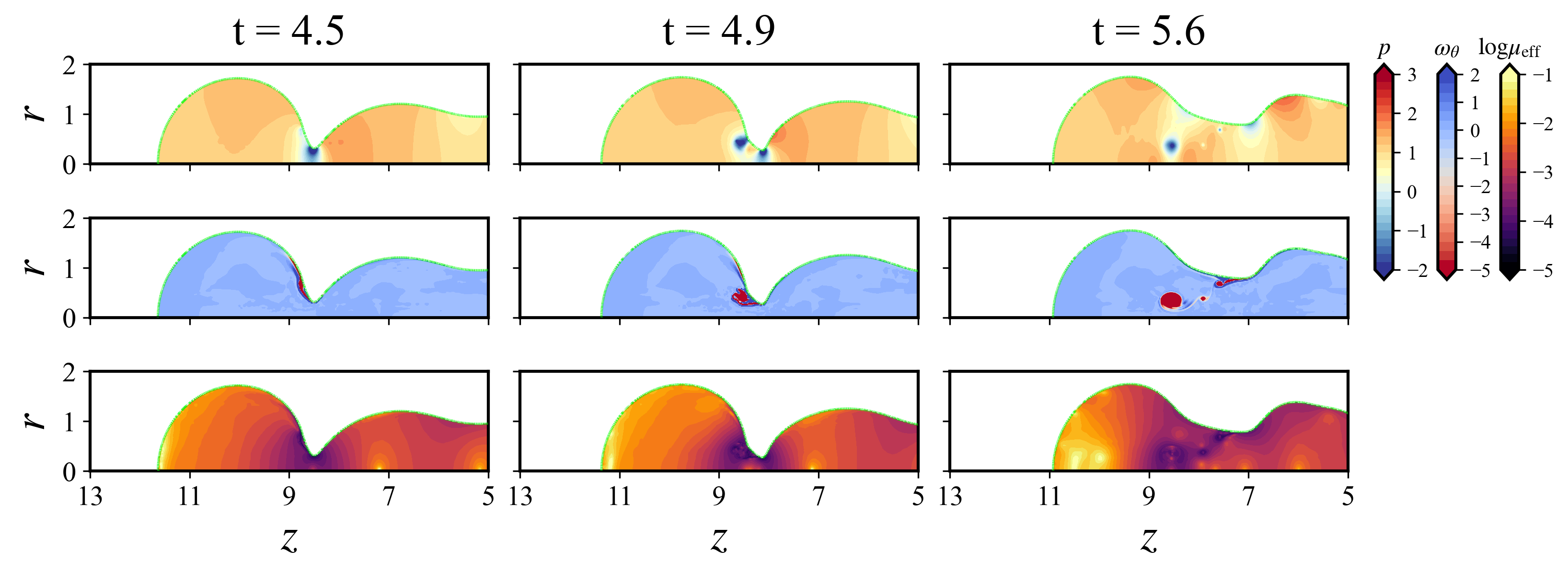}};
  \node[anchor=north west,xshift=-4pt,yshift=-2pt] at (A.north west) {\textbf{(a)}};
\end{tikzpicture}
\begin{tikzpicture}
  \node[inner sep=0] (B) {\includegraphics[width=\linewidth]{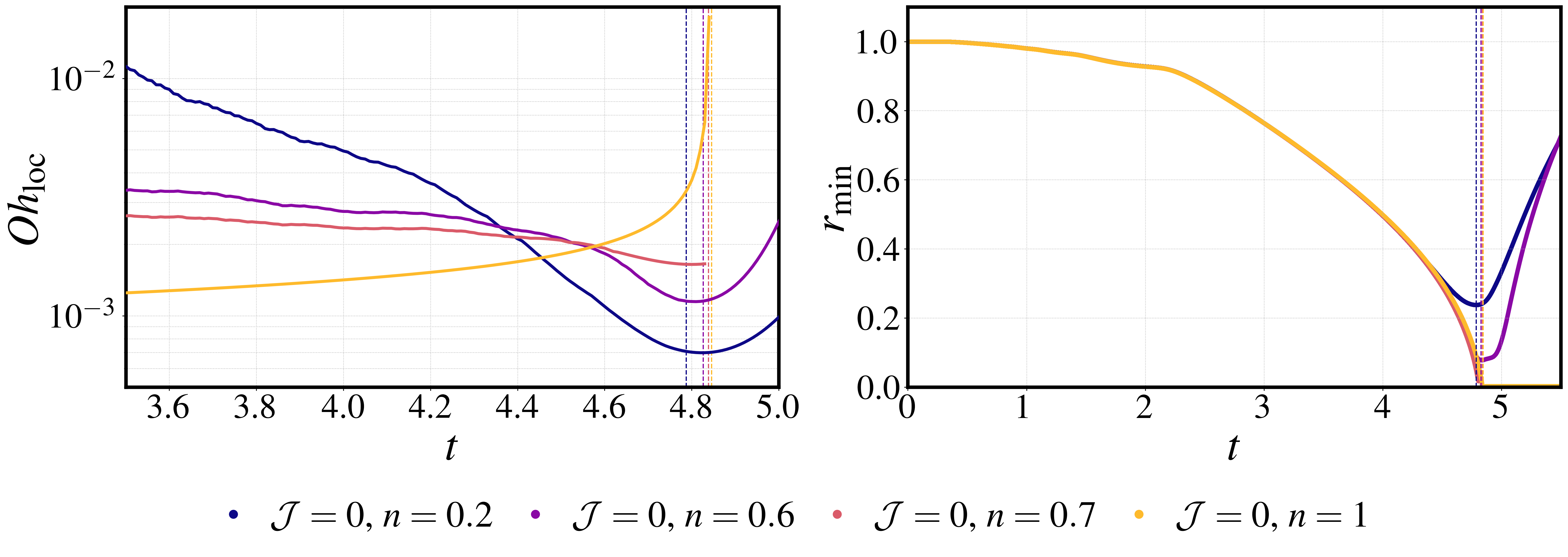}};
  \node[anchor=north west,xshift=-4pt,yshift=2pt] at (B.north west) {\textbf{(b)}};
  \node[anchor=north,xshift=10pt,yshift=2pt] at (B.north) {\textbf{(c)}};
\end{tikzpicture}

\caption{
Shear-thinning reopening regime.
(a) Snapshots of pressure $p$, azimuthal vorticity $\omega_\theta$, and $\log_{10}(\mu_{\mathrm{eff}})$ for the case $Oh_K = 10^{-3}$, $\mathcal{J}=0$, and $n=0.2$, with each row showing one of the three fields at times $t=4.5$, $4.9$, and $5.6$ from left to right.
(b) Temporal evolution of the local Ohnesorge number $Oh_{\text{loc}}$. 
(c)  Minimum neck radius $r_{\min}$ over time for several $n$ at fixed $\mathcal{J} = 0$. The position of the minimum ligament radius, $r_{min}$, is indicated by vertical dashed lines.}
\label{pinchoff_shear_thinning}
\end{figure}

Next, we examine the behaviour of shear-thinning fluids. In the Newtonian limit at low $Oh_K=10^{-3}$, the dynamics are already close to the inertial--capillary regime. One might therefore expect that, for shear-thinning fluids with low flow index $n$, the large strain rates in the neck region reduce the effective viscosity, thereby further weakening viscous resistance and promoting pinch-off.
Unexpectedly,  we find that sufficiently strong shear-thinning at small $\mathcal{J}$ ($n \lesssim 0.6,\,\mathcal{J}\lesssim3.5\times10^{-3}$) can instead lead to reopening (see figure \ref{pinchoff_shear_thinning}). We discuss this in detail below.

The sequence of snapshots in figure \ref{pinchoff_shear_thinning}a shows the evolution of the pressure, azimuthal vorticity, and effective viscosity fields in the vicinity of the neck. 
The structure of the vorticity field differs markedly from the shear-thickening case
as the vorticity is diffuse and does not form a persistent vortex ring, indicating that the reversal of the dynamics is not driven by vorticity generation.
These features coincide with the emergence of a non-uniform pressure field along the axial direction.
The temporal evolution of the local Ohnesorge number, shown in figure \ref{pinchoff_shear_thinning}b, indicates that viscous effects remain spatially heterogeneous and strongly coupled to the evolving strain-rate field. In particular,	$Oh_{\text{loc}}$   decreases during thinning, reflecting the strong reduction in effective viscosity. This indicates that viscous effects weaken relative to inertia as the neck evolves, despite the curvature increase associated with decreasing radius.
The consequences of these dynamics are reflected in figure \ref{pinchoff_shear_thinning}c, where the minimum radius over times departs from  thinning, and undergoes a  reopening. The timing of this reopening correlates with the development of axial gradients in the pressure field.
The mechanism responsible for this transition, involving the interplay between curvature-induced pressure gradients and the spatial redistribution of viscous stresses is analysed in detail below.

To rationalize the shear-thinning reopening, we first revisit mechanisms proposed in the literature when $Oh_K\rightarrow 0$ for Newtonian fluids.
\cite{Wee_2024} showed that escape in two-dimensional sheets  can arise from viscous resistance in the limit $Oh_K \rightarrow 0$, with viscous effects becoming dominant when the minimum sheet thickness   satisfies $h_{\min} \sim Oh_K^{2}$. This scaling implies that the mechanism is only operative once the dynamics reach sufficiently small length scales, typically at $Oh_K \sim \mathcal{O}(10^{-4})$.
\citet{anthony_dynamics_2019} reported end-pinching during the recoiling of liquid threads (free-surface limit)  over the range $Oh_K \in [10^{-4},10^{-3}]$, with only weak sensitivity of the break-up time and neck evolution to  reductions in  $Oh_K$ in this range.  Reopening was also observed for $Oh_K =2\times 10^{-3}$.

\begin{figure}
\centering
\begin{tikzpicture}

\node[inner sep=0] (a) at (0,0)
{\includegraphics[width=0.5\linewidth]{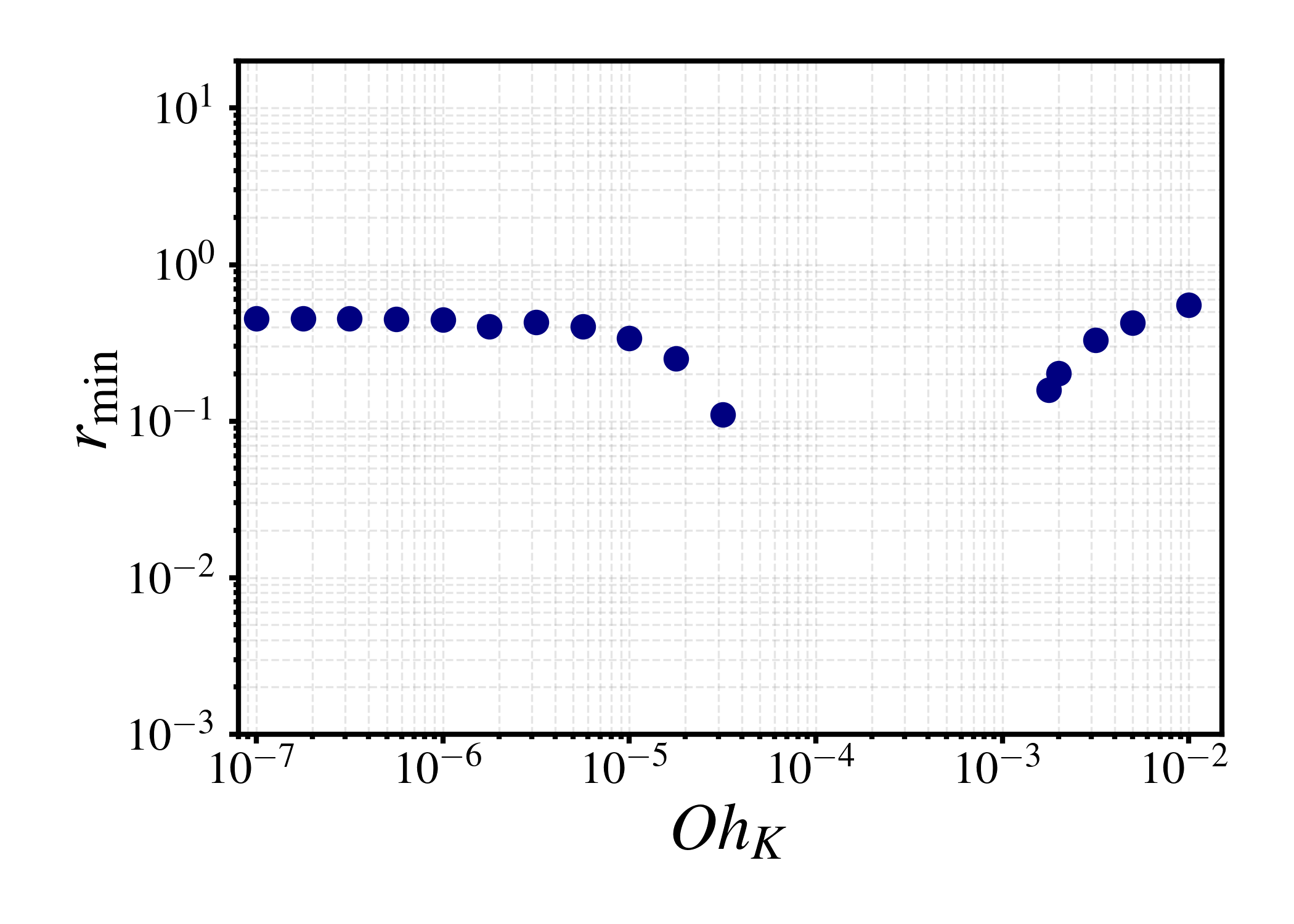}};
\node[anchor=north west, xshift=0pt, yshift=-5pt] at (a.north west)
{\textbf{(a)}};

\node[inner sep=0] (b) at (0.5\linewidth,0)
{\includegraphics[width=0.45\linewidth]{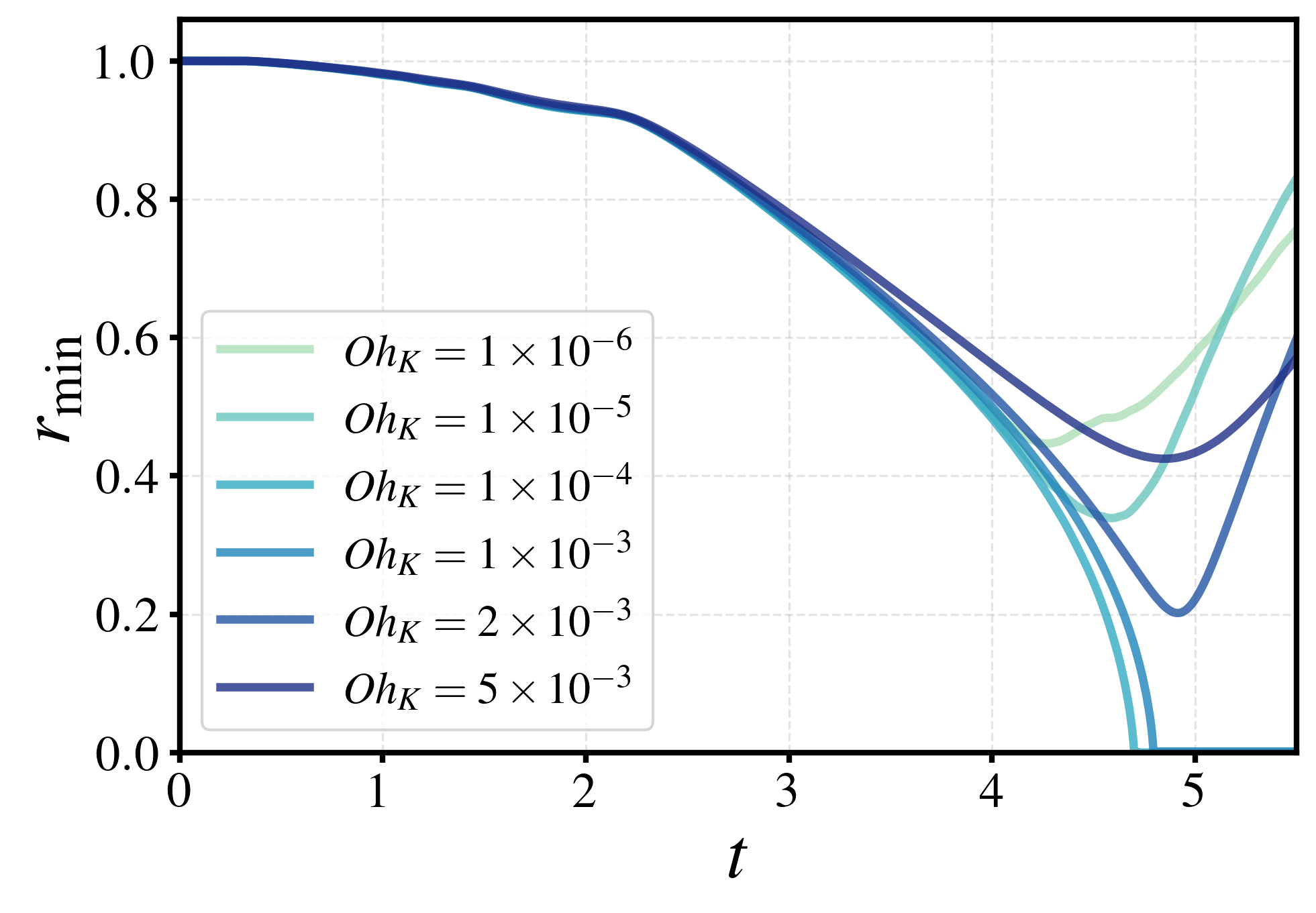}};
\node[anchor=north west, xshift=0pt, yshift=4pt] at (b.north west)
{\textbf{(b)}};

\node[inner sep=0] (c) at (0,-0.35\linewidth)
{\includegraphics[width=0.45\linewidth]{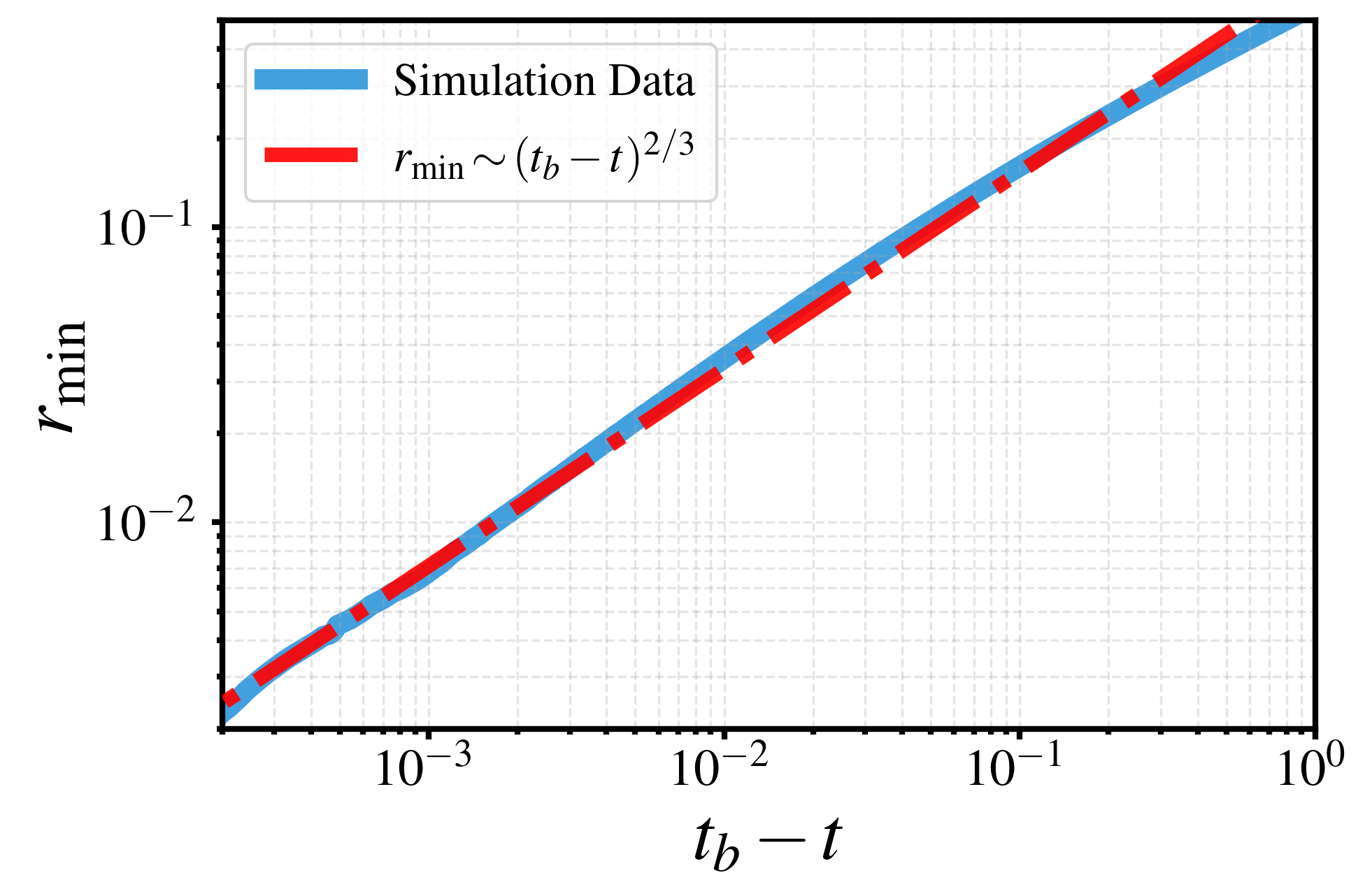}};
\node[anchor=north west, xshift=0pt, yshift=0pt] at (c.north west)
{\textbf{(c)}};

\node[inner sep=0] (d) at (0.5\linewidth,-0.33\linewidth)
{\includegraphics[width=0.35\linewidth]{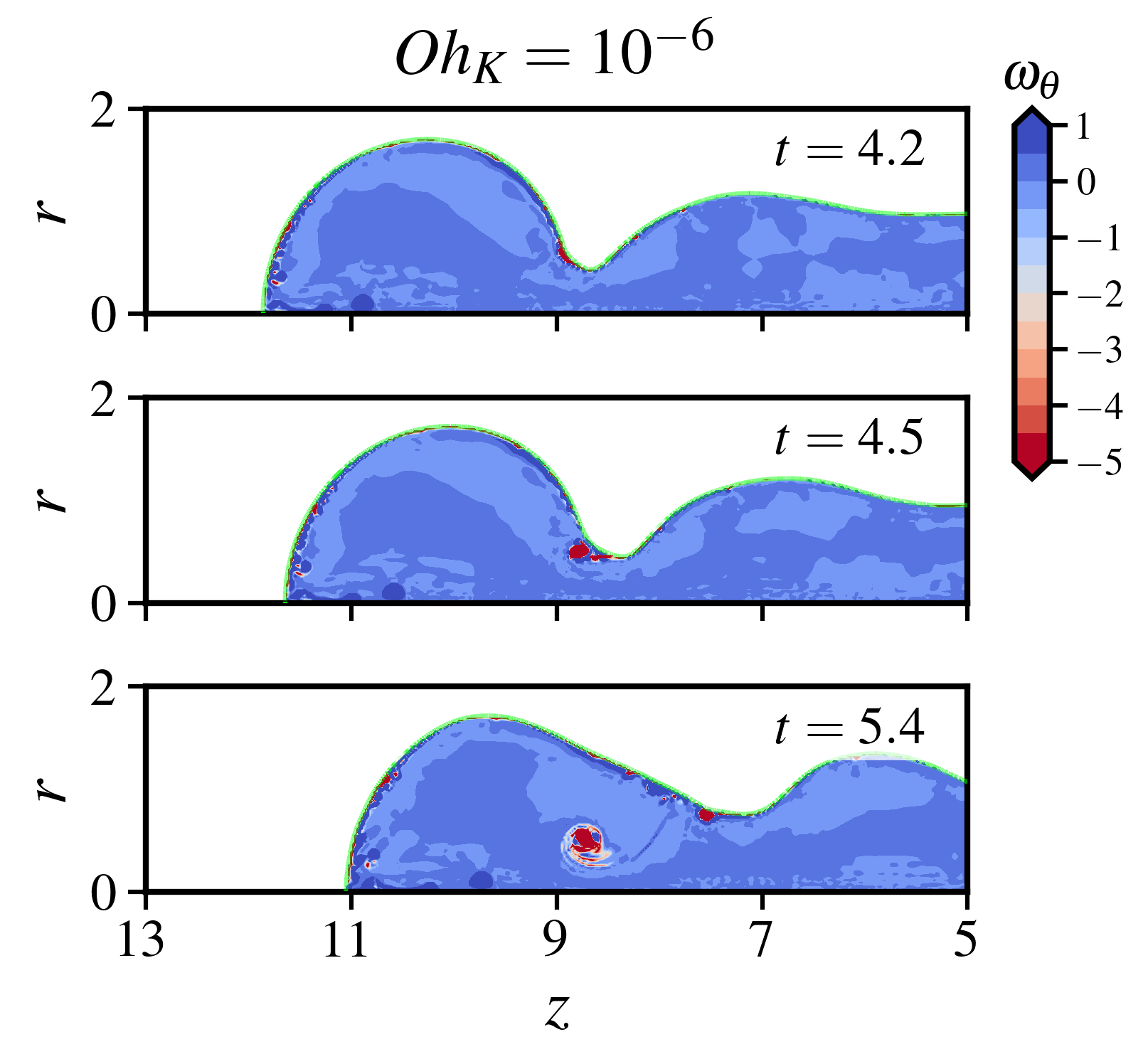}};
\node[anchor=north west, xshift=-10pt, yshift=0pt] at (d.north west)
{\textbf{(d)}};

\end{tikzpicture}

\caption{(a) Minimum radius $r_{\min}$ prior to reopening as a function of Ohnesorge number $Oh_K$. (b) $r_{\min}$ vs $t_b-t$ for a Newtonian ligament at $Oh_K=10^{-3}$, demonstrating the inertial scaling $r_{\min} \!\sim\! (t_b - t)^{2/3}$ {with prefactor $0.712$}. (c) Validation against the viscous-resistance scaling $r_{\min} \sim Oh_K^2$, where $r_{\min}$ denotes the minimum radius $r_{\min}(t)$ at the onset of end-pinching escape.  
(d) Vorticity fields near pinch-off for decreasing Ohnesorge numbers: $Oh_K=10^{-4}$ and $Oh_K=10^{-6}$. Snapshots are shown at successive times before, at, and after the end-pinching escape event. 
}\label{fig:newtonian}
\end{figure}

Figure~\ref{fig:newtonian} presents results for Newtonian threads in the range  $Oh_K\in [10^{-7},10^{-2}]$.
Figure~\ref{fig:newtonian}a shows the evolution of the minimum radius $r_{\min}$ {prior to reopening as a function of $Oh_K$}, where the absence of data indicates cases that undergo pinch-off at the corresponding $Oh_K$. Our simulations are consistent with \citet{anthony_dynamics_2019}: end-pinching break-up is observed for $Oh_K \in [10^{-4},10^{-3}]$, with only weak sensitivity of the break-up time to $Oh_K$ (figure~\ref{fig:newtonian}b). Reopening at $Oh_K = 2\times10^{-3}$ is likewise in agreement with \citet{anthony_dynamics_2019}.
In the break-up cases, the  minimum radius decreases self-similarly, following the classical  inertial-capillary  scaling $r_{\min} \!\sim\! (t_b - t)^{2/3}$ \citep{day_self-similar_1998} (see figure \ref{fig:newtonian}c for $Oh_K=10^{-3}$). The prefactor  is found to be $0.712$ (in agreement with  \citet{kamat_surfactant-driven_2020}).

However, for sufficiently small values, $Oh_K \ll 10^{-5}$, a qualitatively different outcome emerges in which the thread does not pinch-off but instead reopens.
This demonstrates that end-pinching is not the unique inviscid outcome.
In this regime, $r_{\min}$ remains $\mathcal{O}(10^{-1})$ and exhibits a clear plateau for lower $Oh_K$, indicating a reversal of the thinning dynamics.
The flow therefore does not enter the viscously dominated regime identified by \citet{Wee_2024}, and the escape mechanism reported there is not expected to govern the dynamics observed here. 
(Here, using the inertial--capillary scalings $r \sim \tau^{2/3}$, $L_z \sim \tau^{2/3}$, and $u_z \sim \tau^{-1/3}$ in the slender-jet equation yields $\text{inertia} \sim \tau^{-4/3}$ and $\text{viscosity} \sim Oh_K\,\tau^{-5/3}$. Equating these gives $\tau_c \sim Oh_K^3$, and thus $r_{\min} \sim Oh_K^2$, which marks the onset of viscous effects during inertial--capillary thinning.)  Figure~\ref{fig:newtonian}d shows the vorticity field for $Oh_K = 10^{-6}$, revealing a similar behavior as observed in the strongly shear-thinning (see bottom panels of figure \ref{pinchoff_shear_thinning}a). In both cases, the absence of a coherent vortex ring at the neck indicates that reopening does not rely on vortex-mediated viscous resistance, but instead reflects a pressure-driven, inertia-capillary mechanism that persists in the Newtonian limit.

\begin{figure}
\centering
\includegraphics[width=0.9\linewidth]{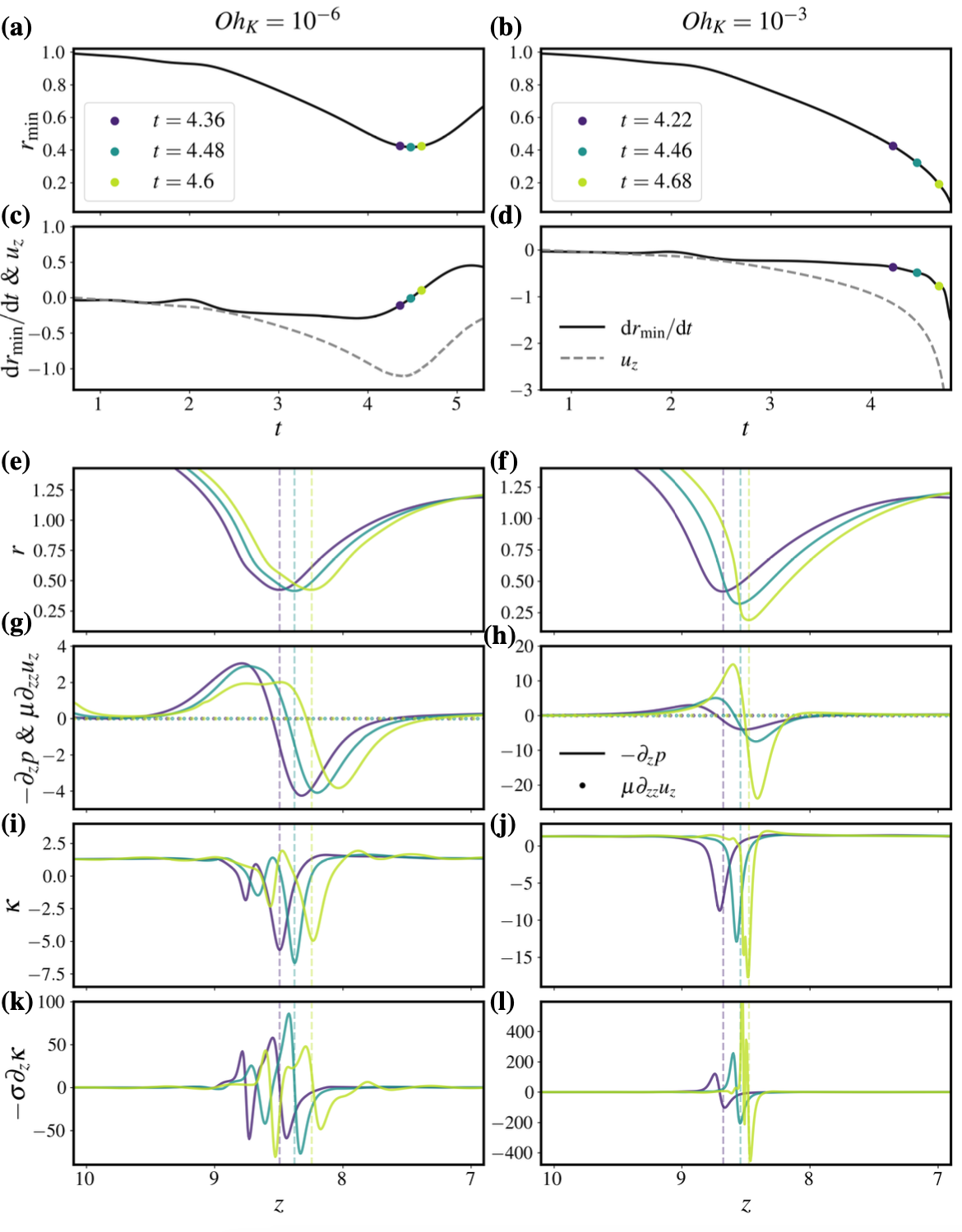}

\caption{Comparison of the dynamics for $Oh_K=10^{-6}$ (left column) and $Oh_K=10^{-3}$ (right column). 
(a,b) Temporal evolution of the minimum radius $r_{\min}$.
(c,d) Temporal evolution of 
$\mathrm{d} r_{\min} / \mathrm{d}t$ and the axial velocity at the neck position, $u_z(z_{\min},t)$.
(e,f) Interface profile $r(z)$ near the neck.
(g,h) Axial momentum balance near the neck, showing $\rho D u_z/D t \approx -\partial_z p$. 
(i,j) Interfacial curvature $\kappa(z)$. 
(k,l)  Capillary forcing $-\sigma \partial_z \kappa$.
Panels (c-d) are colour-coded according to the times shown in panels (a,b). Dashed lines indicate the axial location of the minimum radius.
}
\label{fig:newtonian2}
\end{figure}

As $Oh_K \to 0$, the ligament does not continue toward a singular pinch-off; instead, it undergoes a reopening, with the neck reversing and expanding rather than collapsing. 
This indicates that the apparent break-up convergence reported  in \citet{anthony_dynamics_2019} (see their figure 10)  does not necessarily represent the true asymptotic limit.
We have verified that this behavior is not a numerical artifact. Grid-convergence studies, domain-size independence and time-step refinement yield indistinguishable results. The minimum neck region remains well resolved throughout the evolution, and mass conservation errors remain negligible. In addition, simulations performed in the Newtonian limit ($\varepsilon=0$, $\mathcal{J}=0$) are consistent with an independent purely Newtonian formulation, producing identical interface dynamics and thinning histories (see Appendix C). Simulations reported in figure \ref{fig:newtonian} and \ref{fig:newtonian2} have been performed with level $l=13$.

Next, we will outline the  mechanism responsible for this reopening when $Oh_K \rightarrow 0$, which is governed by curvature-induced pressure gradients that reverse the axial flow near the neck (see figure \ref{fig:newtonian2}). Figure \ref{fig:newtonian2}a,b and \ref{fig:newtonian2}c,d present the temporal evolution of $r_{\min}$, $\mathrm{d}r_{\min}/\mathrm{d}t$, and $u_z(z_{\min},t)$ for the Newtonian cases $Oh_K=10^{-6}$ and $Oh_K=10^{-3}$, corresponding to reopening and pinch-off, respectively.  In the pinch-off case, $\mathrm{d}r_{\min}/\mathrm{d}t$ remains negative and the axial velocity increases in magnitude, indicating sustained capillary-driven collapse. In contrast, for $Oh_K=10^{-6}$, $\mathrm{d}r_{\min}/\mathrm{d}t$ changes sign at finite time, marking a transition to reopening. This transition is accompanied by a loss of axial acceleration at the neck, indicating a breakdown of the collapse mechanism.  

Figure~\ref{fig:newtonian2}e,f shows a magnified view of the interface near the neck during reopening and pinch-off, at three representative times used to elucidate the onset of reopening. Figure \ref{fig:newtonian2}g-h shows the axial momentum balance near the neck. In both cases,
\[
\rho ({D u_z}/{D t}) \approx -\partial_z p,
\]
with negligible viscous contribution (dots in the panels), demonstrating that the dynamics remain inertia--capillary. 
The difference in the pressure distribution during recoil is set by the curvature field through the capillary relation \(p \sim \sigma \kappa\), such that the axial forcing is governed by \(-\partial_z p \approx -\sigma \partial_z \kappa\). 
For \(Oh_K=10^{-3}\), the interface retains a single dominant curvature extremum at the neck, which produces a dipolar pressure-gradient distribution and an  inward acceleration that sustains collapse (see figure~\ref{fig:newtonian2}j,l). 
In contrast, for \(Oh_K=10^{-6}\), inertia--capillary recoil generates a secondary curvature peak upstream of the minimum (see figure~\ref{fig:newtonian2}i). 
At low $Oh_K$, viscous diffusion is too weak to smooth axial variations in the interface shape and velocity field. As a result, axial variations in curvature persist and amplify spatial variations in the capillary pressure gradient, $-\partial_z p \sim -\sigma \partial_z \kappa$. This generates a non-monotonic pressure-gradient field, with neighbouring regions of opposing axial acceleration, which can prevent break-up (see figure~\ref{fig:newtonian2}k). Because viscous diffusion remains weak, these axial variations are not smoothed out. The resulting local force imbalance reverses the neck dynamics, drives fluid away from the minimum, and triggers reopening.

\subsection{No-neck regime}
\label{no-neck-sec}

\begin{figure}
 \includegraphics[width=\linewidth]{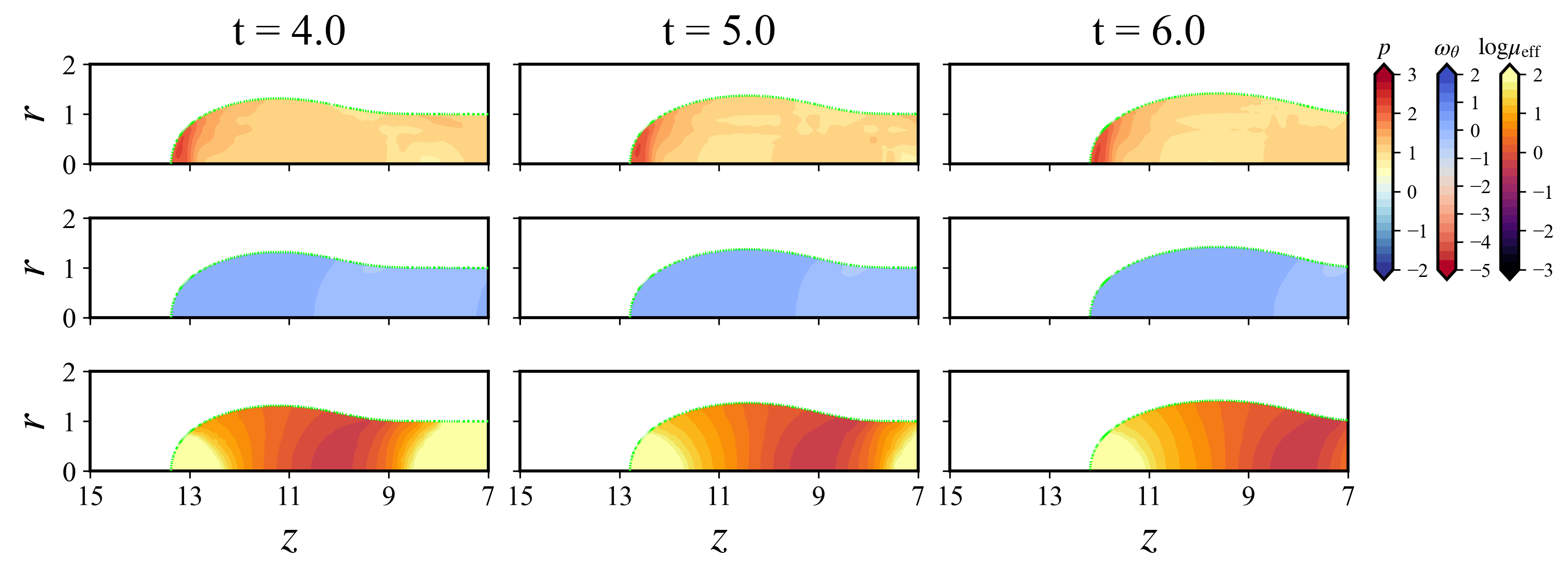}
\caption{No-neck  regime.
 Snapshots of pressure $p$, azimuthal vorticity $\omega_\theta$, and $\log_{10}(\mu_{\mathrm{eff}})$ for the case $Oh_K = 10^{-3}$, $\mathcal{J}=0.3$, and $n=1$, with each row showing one of the three fields at $t = 4.0$, $5.0$, and $6.0$, from left to right. } 
\label{no-neck-vis}
\end{figure}

The no-neck regime corresponds to smooth retraction without neck formation, leading to a single droplet, {as observed in the high-$Oh_K$ limit in Newtonian ligaments due to  viscous effects \citep{PhysRevFluids.5.073602, 10.1063/1.869942}}. The  boundary separating the pinch-off and smooth-retraction regimes arises when the yielded envelope around the neck no longer spans the axial extent of the necked region. To quantify this, we examined the interface profiles from the DNS at the onset of necking. At this stage the ligament is not deformed, thus the minimum radius remains close to  $r_{\min}/R_0\approx1$, while the region over which the interface curvature departs from its far-field value extends axially over $L_z/R_0\simeq2$-$3$.  These values are obtained directly from the simulated profiles and indicate that the capillary disturbance is confined to a few radii near the neck center. 

Capillary forcing at the neck generates an interfacial stress of order
$\tau(a)\sim\sigma/a$.  In axisymmetric creeping flow, the disturbance stress decays approximately as
$\tau(r)\;\sim\;\tau(a)\left(\frac{a}{r}\right)^{2}$,
which follows from the $1/r$ velocity and $1/r^2$ stress fields of the Stokes equations.  The outer radius of the yielded envelope, defined by
$\tau(r_y)=\tau_y$, is therefore
$r_y\;\sim\;\left({\sigma a}/{\tau_y}\right)^{1/2}$.
Pinch-off requires the neck to be fully yielded along its axial extent; once the yield envelope fails to reach the ligament ends ($r_y\lesssim L_z$), the neck remains partially unyielded and can no longer localise.  Setting $r_y\simeq L_z$ gives a threshold
$\tau_y\;\sim\;{\sigma a}/{L_z^{2}}$, and subsequently
$\mathcal J\;\equiv\;{\tau_y R_0}/{\sigma}
\;\sim\;{aR_0}/{L_z^{2}}$,
which defines the onset of the ``no-neck'' regime. Substituting the measured geometric ratios ($a_{\min}/R_0\!\approx\!1$, $L_z/R_0\!\approx\!3$) yields
$ \mathcal J_{\mathrm{end}}=\mathcal{J}/2\;\approx\;
0.0556
$.
This  threshold is in close agreement with the numerically observed transition near
$\mathcal J\simeq0.05$ shown in figure \ref{fig:regime_map}.  The result contains no adjustable parameters and
follows directly from the measured neck geometry and the stress decay implied by the Stokes equations.  Physically, the ends of the ligament, where curvature and capillary stress are largest, remain yielded and retract, while the central neck is shielded by an unyielded core that prevents further localisation and break-up.

\subsection{Motionless regime}
\label{motionless-sec}

\begin{figure}
\includegraphics[width=\linewidth]{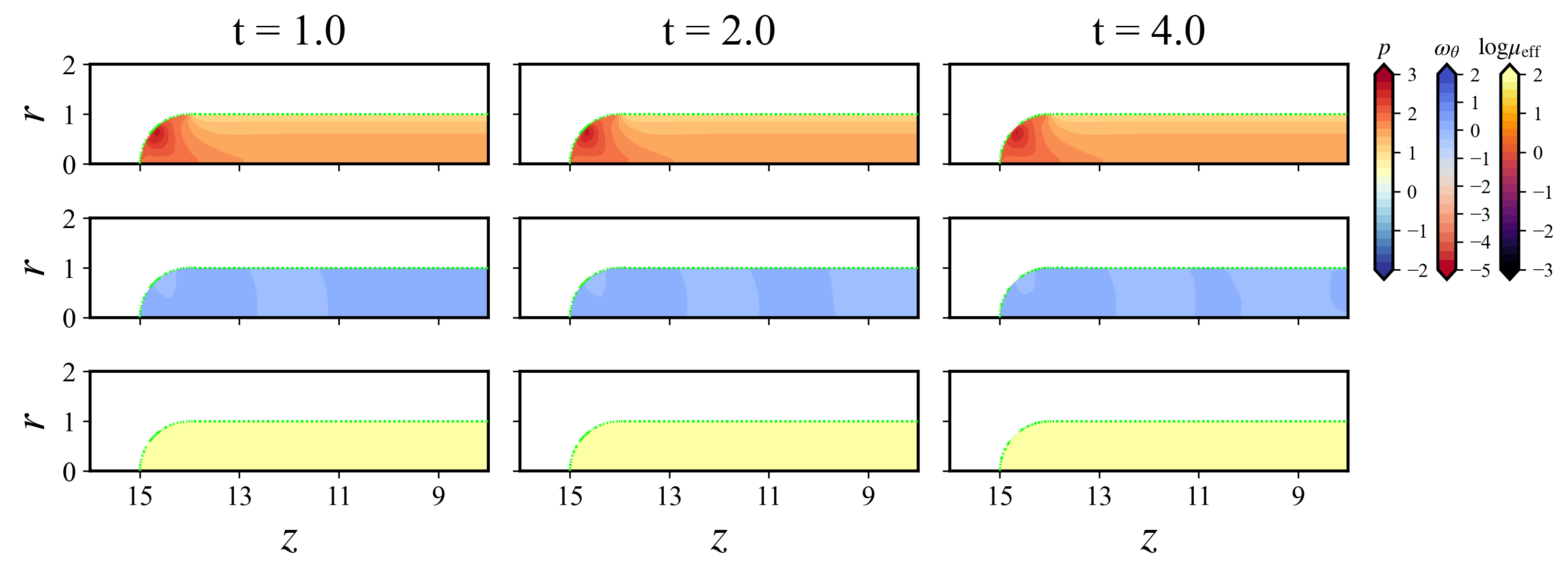} 
\caption{Motionless regime.
 Snapshots of pressure $p$, azimuthal vorticity $\omega_\theta$, and $\log_{10}(\mu_{\mathrm{eff}})$ for the case $Oh_K = 10^{-3}$, $\mathcal{J}=1.0$, and $n=1.0$, with each row showing one of the three fields at   $t =[ 1.0, 2.0, 4.0]$ from left to right.}
\label{fig:Motionless}
\end{figure}

For sufficiently large $\mathcal{J}$, the ligament does not move (see figure \ref{fig:Motionless}).
The boundary separating motionless ligaments from the no-neck regime arises from a simple capillary-yield balance. The associated capillary pressure at the tip of the ligament is $2\sigma/R_0$, while it is  $\sigma/R_0$ at the ligament center. Capillary motion can proceed only if the local pressure difference exceeds the  yield stress $\tau_y$.
When the minimum available capillary stress at the center, $\Delta p_\sigma \simeq \sigma/R_0$, falls below $\tau_y$, the surrounding material remains unyielded and the ligament becomes effectively frozen.
Normalizing stresses by the end-cap value $2\sigma/R_0$ defines the end-cap dimensionless plastocapillary number $\mathcal{J}_{\text{end}} = \tau_y R_0 / (2\sigma)=\mathcal{J}/2$.
Yield-limited arrest of end-cap-driven motion corresponds to  $\tau_y \sim 2\sigma/R_0$, i.e.\ $\mathcal{J}_{\mathrm{end}}=O(1)$.  Accordingly, a numerically observed threshold  $\mathcal{J}_{\mathrm{end}}\simeq 0.5$ corresponds to  $\mathcal{J}\simeq 1$ under the global definition. 
All yield-controlled regime boundaries are interpreted consistently via  $\mathcal{J}_{\mathrm{end}}=\mathcal{J}/2$. 
Since this criterion involves only stresses and not strain rates, the flow-behavior index $n$ plays no role in setting the arrest threshold. This vertical boundary in the $(\mathcal{J},n)$ plane is therefore fully determined by geometric curvature asymmetry and is independent of the rheological index~$n$. This agrees well with the predictions of the simulations.

\section{Conclusions}
\label{conclude-sec}

Capillary retraction of liquid ligaments is well understood for Newtonian fluids, whereas the role of yield stress remains comparatively unexplored. Here, we consider Herschel-Bulkley fluids, which incorporate both a finite yield stress and a nonlinear, shear-rate-dependent viscosity. Unlike Bingham models, the Herschel-Bulkley constitutive law yields a continuous, strain-rate-dependent viscosity.
Direct numerical simulations using \textsc{Basilisk~C} identify four distinct dynamical outcomes---pinch-off, escape, no-neck formation, and complete arrest---whose occurrence depends on the coupled roles of the flow-behaviour index $n$ and the plastocapillary number $\mathcal{J}$. The resulting regime map is constructed in the $(n,\mathcal{J})$ plane for a viscoplatic ligament with $Oh_K=10^{-3}$ retracting in air.
The choice $Oh_K=10^{-3}$
is consistent with prior studies of inertial-capillary ligament retraction \citep{notz_dynamics_2004,kamat_surfactant-driven_2020}. This regime is characterized by end-pinching in the Newtonian case and provides a baseline for direct comparison, allowing the effects of yield stress and shear-dependent viscosity to be assessed relative to inertial-capillary dynamics.

The shear-dependent viscosity gives rise to mechanisms that are absent in Bingham fluids, including (i) escape from pinch-off via vorticity generation in shear-thickening regimes, and (ii) escape from pinch-off via curvature-induced flow, associated with non-uniform axial pressure gradients, in shear-thinning regimes.
The shear-thinning mechanism persists in the Newtonian limit of vanishing viscosity, where it becomes a purely inertial-capillary pathway for reopening that, to the best of our knowledge, has not been reported in studies of ligament retraction. This mechanism provides a route to suppress end-pinching that cannot be captured by constant-viscosity yield-stress models (i.e., Bingham models).
Finally,  for moderately large $\mathcal{J}$, capillary stresses are sufficient to induce limited deformation but remain too weak to 
initiate necking. The ligament therefore deforms without forming a neck, leading to a no-neck regime.
As $\mathcal{J}$ is increased further, the characteristic capillary stress falls entirely below the yield threshold, and the material remains unyielded throughout. In this limit, capillarity is unable to drive any flow, and the ligament remains fully arrested.
Together, these results demonstrate that the fate of a retracting viscoplastic ligament is governed by the interplay of capillarity, viscous dissipation, shear-dependent rheology, and yield stress, and therefore cannot be captured within the Bingham limit.

Previous free-surface numerical studies reported end-pinching in Newtonian ligaments down to $Oh_K \sim 10^{-4}$, with break-up times and neck evolution exhibiting only weak sensitivity to further reductions in $Oh_K$ \citep[see figure 10 of][]{anthony_dynamics_2019}. Based on this apparent convergence, they interpreted the dynamics in this range as effectively inviscid. However, this conclusion is drawn over a limited interval, $Oh_K \in [10^{-4}-10^{-3}]$.
While we recover this behaviour over the same range, extending the simulations to significantly lower 
$Oh_K$	 reveals a qualitatively different regime in which, instead of pinch-off, the ligament undergoes inertial-capillary reopening.
This behaviour, not previously reported in the context of Newtonian ligaments, is also observed in strongly shear-thinning fluids, where the effective viscosity in the neck region is reduced. The present results therefore establish a continuous connection between strongly shear-thinning dynamics and the Newtonian limit as $Oh_K \to 0$, both governed by an inertial--capillary balance in which viscous stresses are insufficient to sustain singular thinning. We note that existing experiments on liquid threads have typically been restricted to $Oh_K \gtrsim 10^{-3}$ \citep{kamat_surfactant-driven_2020,castrejon-pita_breakup_2012}, and reaching the regime $Oh_K \lesssim 10^{-5}$ may be challenging in practice. Nevertheless, the present results demonstrate that the asymptotic limit is not characterized by  pinch-off, 
but instead admits  reopening as an alternative outcome. Finally, we have verified that this behaviour is robust and not a numerical artifact.

We acknowledge that  most yield-stress fluids encountered experimentally, such as Carbopol gels, exhibit measurable viscoelasticity in addition to their plastic response \citep{tanver,franca2026coalescenceprintedyieldstress}.  The present study isolates the purely viscoplastic limit through the Herschel-Bulkley formulation, thereby clarifying the individual roles of yield stress and shear-dependent viscosity in shaping end-pinching escape.  While this simplification provides a useful theoretical baseline, the resulting dynamics cannot be mapped directly onto real yield-stress materials. Future work should therefore examine how elastic stress storage and relaxation modify the reopening mechanism 
identified here.

\medskip

\subsection*{Acknowledgments }
CRCA thanks Sascha Hilgenfeldt for helpful comments on an earlier version of the manuscript. 
This research used the Delta advanced computing and data resource which is supported by the National Science Foundation (award OAC 2005572) and the State of Illinois.

\section*{Appendix A: Sensitivity to the regularization parameter  }

The Herschel-Bulkley model is implemented using a regularized formulation in which the apparent viscosity is smoothed through an $\varepsilon$-dependent function to prevent singular behaviour as the strain-rate magnitude $||\boldsymbol{\mathcal{D}}|| \to 0$. As a result, the yield surface is not sharply defined, since $||\boldsymbol{\mathcal{D}}||$ remains finite for any $\varepsilon > 0$, and the effective viscosity is bounded by $\mu_{\mathrm{eff}} \sim \tau_y / \varepsilon$ in weakly deformed regions.

\begin{figure}
    \centering
    \begin{tikzpicture}
        \node[inner sep=0] (fig) {\includegraphics[width=0.9\linewidth]{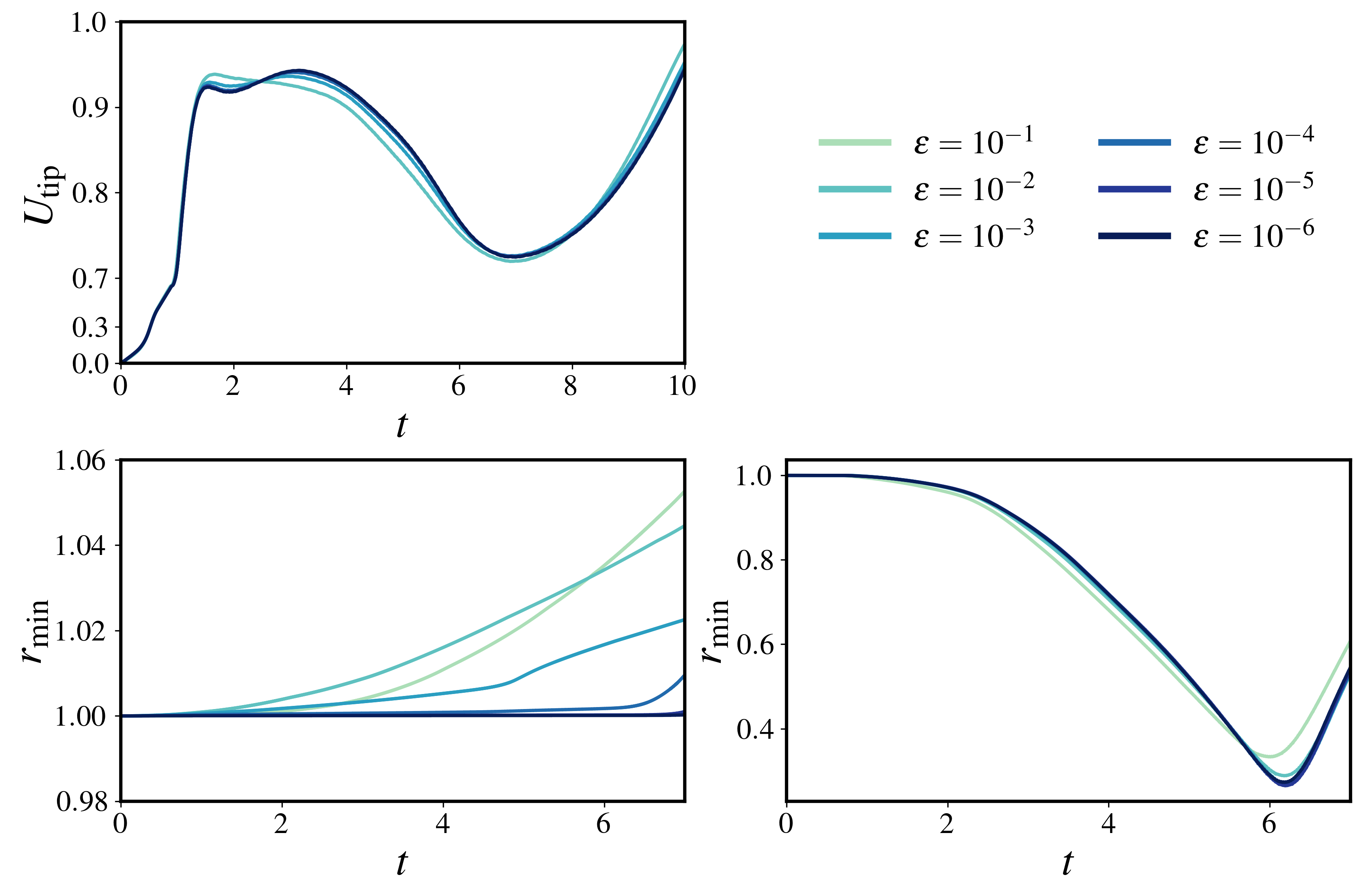}};
    
        \node[anchor=north west, xshift=0pt, yshift=0pt] at (fig.north west) {\textbf{(a)}};

        \node[anchor=north west, xshift=-4pt, yshift=10pt] at (fig.west) {\textbf{(b)}};
        
        \node[anchor=center, xshift=8pt, yshift=4pt] at (fig.center) {\textbf{(c)}};
    \end{tikzpicture} 
\caption{Verification of $\varepsilon$-independence for $\varepsilon \in [10^{-6},\,10^{-1}]$.
(a) Ligament tip velocity $U_{\mathrm{tip}}$ for $Oh_K = 0.005$, $\mathcal{J} = 0.05$, $n = 2$.
(b) Temporal evolution of the minimum radius $r_{\min}$ in the motionless regime ($\mathcal{J} = 0.5$, $n = 1$).
(c) Temporal evolution of $r_{\min}$ for a case near the transition between end-pinching and pinch-off ($\mathcal{J} = 0.03$, $n = 1.125$).}
    \label{fig:epsilon}
\end{figure}

To assess the influence of the regularization parameter, figure~\ref{fig:epsilon}a shows the evolution of the tip velocity $U_{\mathrm{tip}}(t)$ for a retracting ligament at $Oh_K = 0.005$, $\mathcal{J} = 0.05$, and $n = 2$, for $\varepsilon$ varying from $10^{-2}$ to $10^{-7}$. While small differences are observed at early times for larger $\varepsilon$, all curves collapse for $\varepsilon \leq 10^{-4}$, indicating that the dynamics are insensitive to further reductions of the regularization parameter. 

We next examine the sensitivity of the motionless regime to the regularization. This regime corresponds to a near-arrest of the dynamics, where $||\boldsymbol{\mathcal{D}}|| \to 0$ and the regularization has the strongest potential influence through the viscosity cap. As shown in figure~\ref{fig:epsilon}b for $\mathcal{J}=0.5$ and $n=1$, the minimum neck radius exhibits a plateau with variations below $1\%$ over $\varepsilon \in [10^{-4},10^{-6}]$. This demonstrates that the motionless regime is not induced by the regularization, but is a genuine physical feature of the flow. Finally, we assess the robustness of the transition between end-pinching and pinch-off near the regime boundary. Figure~\ref{fig:epsilon}c shows the evolution of $r_{\min}(t)$ for a representative case ($\mathcal{J}=0.03$, $n=1.125$) with $\varepsilon \in [10^{-1},10^{-6}]$. All cases exhibit end-pinching, and for sufficiently small values $\varepsilon \leq 10^{-4}$, the curves collapse with relative discrepancies below $0.1\%$ in $r_{\min}$. This confirms quantitative convergence of the dynamics and demonstrates that the location of the regime boundary is not an artifact of the regularization.

Based on  these checks, we have adopted $\varepsilon = 10^{-4}$ or lower  for all our simulations.

\section*{Appendix B: Grid dependence study  }

\begin{figure}
    \centering
    \includegraphics[width=0.5\linewidth]{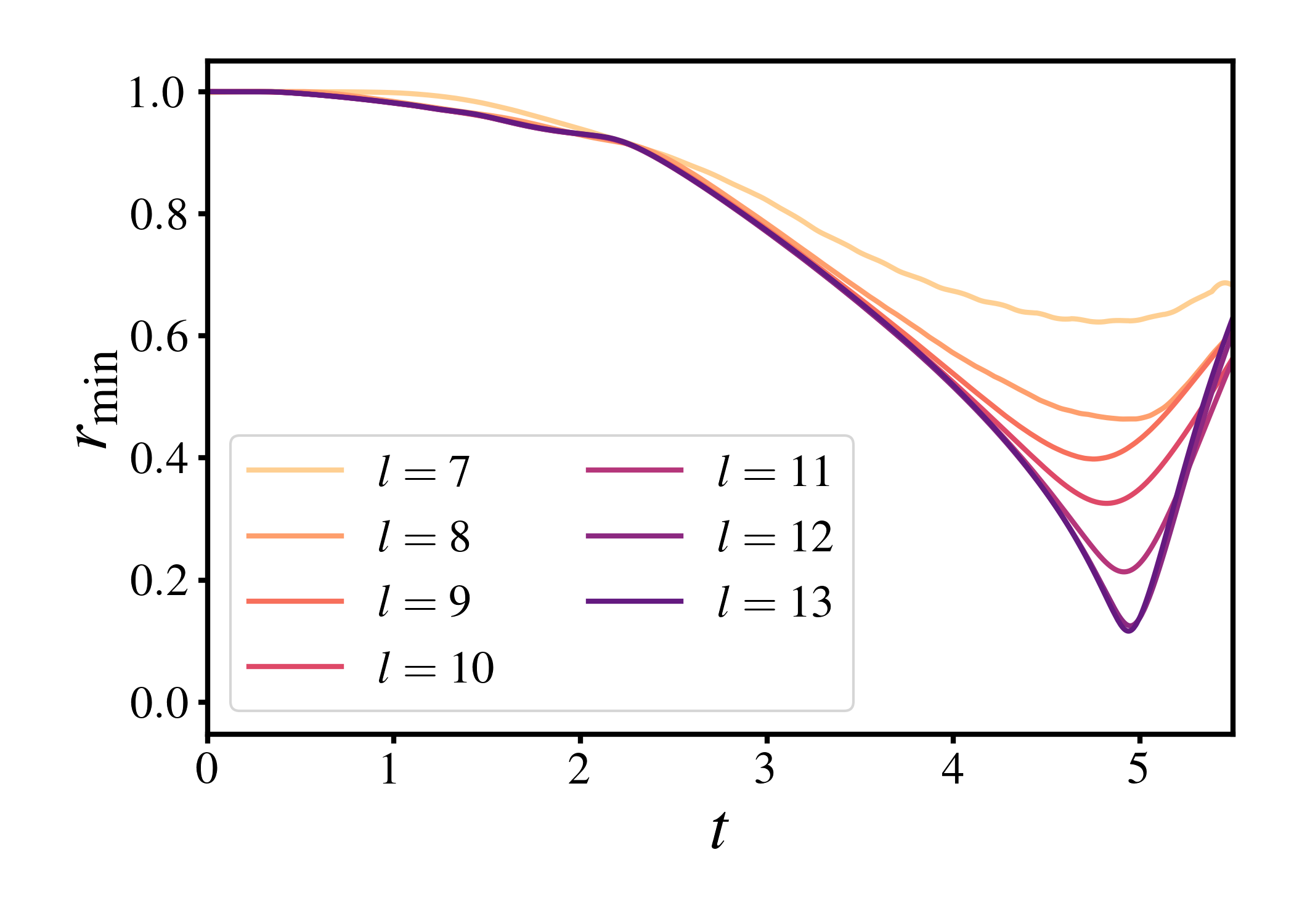}
\caption{Mesh-dependence study using different maximum grid levels, $l = 7$--$13$. Temporal evolution of the minimum radius $r_{\min}$ for a case near the transition between end-pinching and pinch-off, with $Oh_K = 10^{-3}$, $\mathcal{J} = 0.002$, and $n = 1.125$.}
\label{fig:lvl_hmin}
\end{figure}

A mesh-dependence study was performed by varying the maximum grid level from $l=7$ to $l=13$ at fixed $Oh_K=10^{-3}$, $\mathcal{J}=0.002$, and $n=1.125$. This parameter set corresponds to a case near the boundary between break-up and reopening in the $(\mathcal{J},n)$ regime diagram.
As shown in figure~\ref{fig:lvl_hmin}, the solutions exhibit clear convergence with increasing resolution. In particular, the results for $l=12$ and $l=13$ are nearly indistinguishable over the entire evolution, including the minimum radius and the subsequent reopening, indicating that the solution is effectively grid-converged at these resolutions.  These results demonstrate that a maximum grid level $l=12$ is sufficient to capture both the thinning and reopening dynamics with quantitative accuracy. For cases approaching the limit $Oh_K \rightarrow 0$, all simulations were checked using a maximum grid level $l=13$ for both Newtonian and non-Newtonian fluids. This level of refinement is necessary to ensure that  the strong curvature gradients that arise in the inertia-capillary regime are well resolved as they  govern the reopening dynamics of the ligament.

\section*{Appendix C: Validation of the solver in the effectively inviscid limit, $Oh_K \rightarrow0$}
\label{append:C}

\begin{figure}
    \centering
 \includegraphics[width=0.5\linewidth]{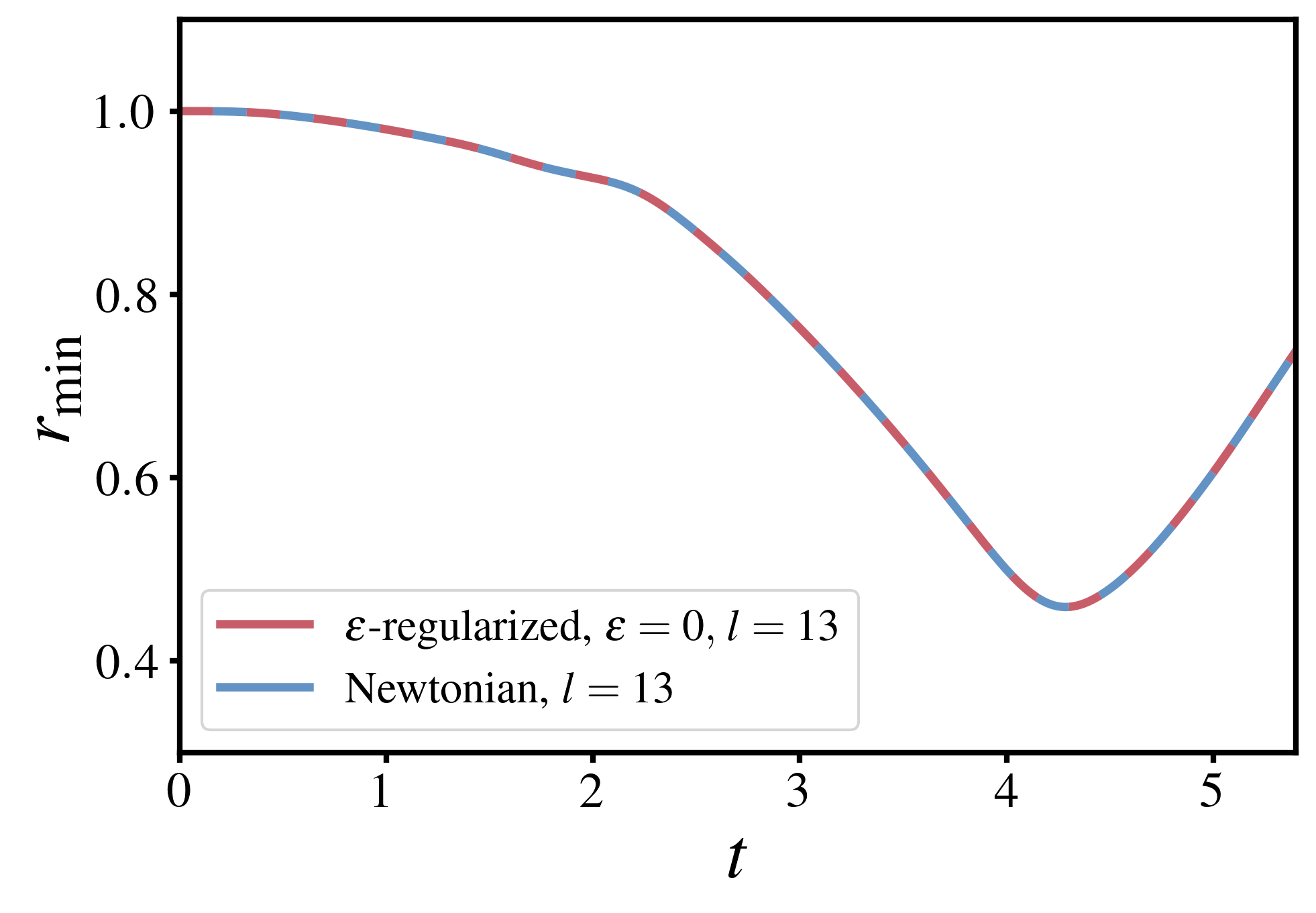}
    \caption{{Comparison of the interface evolution between the original Newtonian solver  and the $\varepsilon$-regularized formulation  at $Oh_K = 10^{-6}$. Both computations are performed at a maximum grid level $l=13$. 
    }}
    \label{fig:Compare_VP_Nt}
\end{figure}

The solver is based on $\varepsilon$-regularized Herschel-Bulkley constitutive relations. In addition to the validation against \citet{notz_dynamics_2004} provided in figure \ref{fig:computational_domain}b, we further assess the formulation in the nearly inviscid limit, where $Oh_K \sim 10^{-6}$. In figure \ref{fig:Compare_VP_Nt}, the current solver is compared with a standard Newtonian formulation \citep{popinet2009accurate,BasiliskPlateauTest}.
The excellent agreement between the two solutions at  $Oh_K \sim 10^{-6}$  shows that the regularized formulation recovers the Newtonian dynamics in the effectively inviscid regime, and that the observed  escape mechanism is not an artifact of regularization.

{\section*{Appendix D: Validation of Herschel-Bulkley model }}

\begin{figure}
\centering
\includegraphics[width=0.7\linewidth]{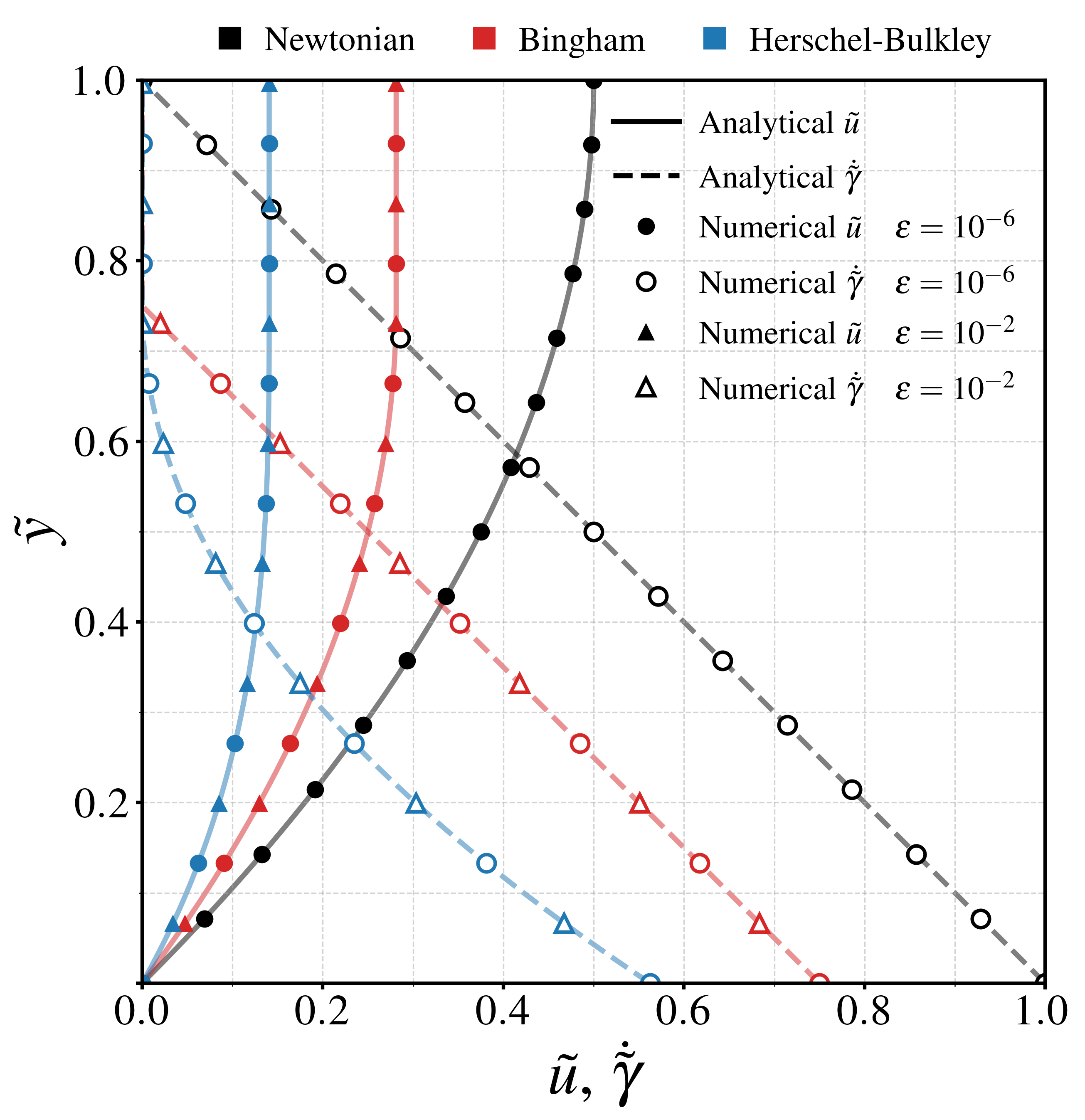}%
\caption{Validation of the $\varepsilon$-regularized formulation using steady pressure-driven planar Poiseuille flow.
Solid and dashed lines denote the analytical solutions of the velocity $\tilde{u}$ and shear-rate $\dot{\tilde{\gamma}}$ profiles as functions of the wall-normal coordinate $\tilde{y}$,
while circular and triangular markers represent the numerical results obtained with $\varepsilon = 10^{-6}$ and $10^{-2}$, respectively.}
\label{fig:HB_validation}
\end{figure}

To validate the $\varepsilon$-regularized viscoplastic formulation, we consider steady pressure-driven planar Poiseuille flow of a Herschel-Bulkley fluid, for which an analytical solution is available \citep{panaseti2018pressure}. In the Stokes limit, the shear stress varies linearly across the channel as $\tilde{\mathcal T}(\tilde y)=1-\tilde y$, yielding a plug region for $\tilde y \ge \tilde Y=1-\tilde{\mathcal T}_y$. In the yielded region ($\tilde y<\tilde Y$), the constitutive relation gives the shear-rate
\begin{equation}
\tilde{\dot\gamma}(\tilde y)=\left(\frac{\tilde Y-\tilde y}{\tilde K}\right)^{1/n},
\end{equation}
while $\tilde{\dot\gamma}=0$ in the plug. Integration with no-slip at the wall yields the velocity profile
\begin{equation}
\tilde u(\tilde y)=\frac{n}{n+1}\tilde K^{-1/n}
\left[
\tilde Y^{\frac{n+1}{n}}-(\tilde Y-\tilde y)^{\frac{n+1}{n}}
\right], \quad \tilde y<\tilde Y,
\end{equation}
with a constant velocity in the plug region. This canonical solution provides a stringent benchmark for assessing the accuracy of the regularized formulation, particularly its ability to capture the yield surface location and the transition between yielded and unyielded regions.

Figure~\ref{fig:HB_validation} presents a direct comparison between the numerical solutions obtained using the $\varepsilon$-regularized constitutive law and the exact analytical solutions for steady, fully developed channel flow of Newtonian ($\tilde{\mu} = 1.0, \tilde{\mathcal{T}}_y=0$), Bingham $(\tilde{K} = 1.0, \tilde{\mathcal{T}}_y = 0.25, n=1.0)$, and Herschel-Bulkley $(\tilde{K} = 1.0, \tilde{\mathcal{T}}_y = 0.25, n=0.5)$ fluids. 
The comparison includes both the velocity profile $\tilde{u}(\tilde{y})$ and the shear-rate distribution $\tilde{\dot{\gamma}}(\tilde{y})$ across the  channel height, encompassing both yielded regions and the central plug where applicable. For the Newtonian case, the numerical solution recovers the exact parabolic velocity profile and linear shear-rate distribution dictated by the pressure-viscous balance. In the Bingham and Herschel-Bulkley cases, the computations reproduce the finite plug region, the analytically predicted plug width, and the piecewise structure of the solution. The shear-rate vanishes within the unyielded core and exhibits a sharp transition at the yield surface, consistent with the theoretical stress distribution. We observe that the solutions are effectively independent of the regularization parameter, as the numerical results are indistinguishable for $\varepsilon \in [10^{-6},10^{-2}]$ when compared with the analytical predictions. These findings confirm that the regularized formulation used in this work consistently recovers the exact Newtonian limit and the analytical viscoplastic solutions in both yielded and unyielded regions.

\bibliography{references}

\begin{thebibliography}{55}%
\makeatletter
\providecommand \@ifxundefined [1]{%
 \@ifx{#1\undefined}
}%
\providecommand \@ifnum [1]{%
 \ifnum #1\expandafter \@firstoftwo
 \else \expandafter \@secondoftwo
 \fi
}%
\providecommand \@ifx [1]{%
 \ifx #1\expandafter \@firstoftwo
 \else \expandafter \@secondoftwo
 \fi
}%
\providecommand \natexlab [1]{#1}%
\providecommand \enquote  [1]{``#1''}%
\providecommand \bibnamefont  [1]{#1}%
\providecommand \bibfnamefont [1]{#1}%
\providecommand \citenamefont [1]{#1}%
\providecommand \href@noop [0]{\@secondoftwo}%
\providecommand \href [0]{\begingroup \@sanitize@url \@href}%
\providecommand \@href[1]{\@@startlink{#1}\@@href}%
\providecommand \@@href[1]{\endgroup#1\@@endlink}%
\providecommand \@sanitize@url [0]{\catcode `\\12\catcode `\$12\catcode `\&12\catcode `\#12\catcode `\^12\catcode `\_12\catcode `\%12\relax}%
\providecommand \@@startlink[1]{}%
\providecommand \@@endlink[0]{}%
\providecommand \url  [0]{\begingroup\@sanitize@url \@url }%
\providecommand \@url [1]{\endgroup\@href {#1}{\urlprefix }}%
\providecommand \urlprefix  [0]{URL }%
\providecommand \Eprint [0]{\href }%
\providecommand \doibase [0]{https://doi.org/}%
\providecommand \selectlanguage [0]{\@gobble}%
\providecommand \bibinfo  [0]{\@secondoftwo}%
\providecommand \bibfield  [0]{\@secondoftwo}%
\providecommand \translation [1]{[#1]}%
\providecommand \BibitemOpen [0]{}%
\providecommand \bibitemStop [0]{}%
\providecommand \bibitemNoStop [0]{.\EOS\space}%
\providecommand \EOS [0]{\spacefactor3000\relax}%
\providecommand \BibitemShut  [1]{\csname bibitem#1\endcsname}%
\let\auto@bib@innerbib\@empty
\bibitem [{\citenamefont {Anthony}\ \emph {et~al.}(2019)\citenamefont {Anthony}, \citenamefont {Kamat}, \citenamefont {Harris},\ and\ \citenamefont {Basaran}}]{anthony_dynamics_2019}%
  \BibitemOpen
  \bibfield  {author} {\bibinfo {author} {\bibfnamefont {C.~R.}\ \bibnamefont {Anthony}}, \bibinfo {author} {\bibfnamefont {P.~M.}\ \bibnamefont {Kamat}}, \bibinfo {author} {\bibfnamefont {M.~T.}\ \bibnamefont {Harris}},\ and\ \bibinfo {author} {\bibfnamefont {O.~A.}\ \bibnamefont {Basaran}},\ }\bibfield  {title} {\bibinfo {title} {Dynamics of contracting filaments},\ }\href {https://doi.org/10.1103/PhysRevFluids.4.093601} {\bibfield  {journal} {\bibinfo  {journal} {Phys. Rev. Fluids}\ }\textbf {\bibinfo {volume} {4}},\ \bibinfo {pages} {093601} (\bibinfo {year} {2019})}\BibitemShut {NoStop}%
\bibitem [{\citenamefont {Schulkes}(1996)}]{schulkes_contraction_1996}%
  \BibitemOpen
  \bibfield  {author} {\bibinfo {author} {\bibfnamefont {R.~M. S.~M.}\ \bibnamefont {Schulkes}},\ }\bibfield  {title} {{\selectlanguage {en}\bibinfo {title} {The contraction of liquid filaments}},\ }\href {https://doi.org/10.1017/S0022112096001632} {\bibfield  {journal} {\bibinfo  {journal} {J. Fluid Mech.}\ }\textbf {\bibinfo {volume} {309}},\ \bibinfo {pages} {277} (\bibinfo {year} {1996})}\BibitemShut {NoStop}%
\bibitem [{\citenamefont {Anthony}\ \emph {et~al.}(2023)\citenamefont {Anthony}, \citenamefont {Wee}, \citenamefont {Garg}, \citenamefont {Thete}, \citenamefont {Kamat}, \citenamefont {Wagoner}, \citenamefont {Wilkes}, \citenamefont {Notz}, \citenamefont {Chen}, \citenamefont {Suryo} \emph {et~al.}}]{anthony2023sharp}%
  \BibitemOpen
  \bibfield  {author} {\bibinfo {author} {\bibfnamefont {C.~R.}\ \bibnamefont {Anthony}}, \bibinfo {author} {\bibfnamefont {H.}~\bibnamefont {Wee}}, \bibinfo {author} {\bibfnamefont {V.}~\bibnamefont {Garg}}, \bibinfo {author} {\bibfnamefont {S.~S.}\ \bibnamefont {Thete}}, \bibinfo {author} {\bibfnamefont {P.~M.}\ \bibnamefont {Kamat}}, \bibinfo {author} {\bibfnamefont {B.~W.}\ \bibnamefont {Wagoner}}, \bibinfo {author} {\bibfnamefont {E.~D.}\ \bibnamefont {Wilkes}}, \bibinfo {author} {\bibfnamefont {P.~K.}\ \bibnamefont {Notz}}, \bibinfo {author} {\bibfnamefont {A.~U.}\ \bibnamefont {Chen}}, \bibinfo {author} {\bibfnamefont {R.}~\bibnamefont {Suryo}}, \emph {et~al.},\ }\bibfield  {title} {\bibinfo {title} {Sharp interface methods for simulation and analysis of free surface flows with singularities: breakup and coalescence},\ }\href@noop {} {\bibfield  {journal} {\bibinfo  {journal} {Annu. Rev. Fluid Mech.}\ }\textbf {\bibinfo {volume} {55}},\ \bibinfo {pages} {707} (\bibinfo {year} {2023})}\BibitemShut {NoStop}%
\bibitem [{\citenamefont {Lohse}(2022)}]{lohse2022fundamental}%
  \BibitemOpen
  \bibfield  {author} {\bibinfo {author} {\bibfnamefont {D.}~\bibnamefont {Lohse}},\ }\bibfield  {title} {\bibinfo {title} {Fundamental fluid dynamics challenges in inkjet printing},\ }\href@noop {} {\bibfield  {journal} {\bibinfo  {journal} {Annu. Rev. Fluid Mech.}\ }\textbf {\bibinfo {volume} {54}},\ \bibinfo {pages} {349} (\bibinfo {year} {2022})}\BibitemShut {NoStop}%
\bibitem [{\citenamefont {Castrej\'on-Pita}\ \emph {et~al.}(2012)\citenamefont {Castrej\'on-Pita}, \citenamefont {Castrej\'on-Pita},\ and\ \citenamefont {Hutchings}}]{PRL_ACP}%
  \BibitemOpen
  \bibfield  {author} {\bibinfo {author} {\bibfnamefont {A.~A.}\ \bibnamefont {Castrej\'on-Pita}}, \bibinfo {author} {\bibfnamefont {J.~R.}\ \bibnamefont {Castrej\'on-Pita}},\ and\ \bibinfo {author} {\bibfnamefont {I.~M.}\ \bibnamefont {Hutchings}},\ }\bibfield  {title} {\bibinfo {title} {Breakup of liquid filaments},\ }\href {https://doi.org/10.1103/PhysRevLett.108.074506} {\bibfield  {journal} {\bibinfo  {journal} {Phys. Rev. Lett.}\ }\textbf {\bibinfo {volume} {108}},\ \bibinfo {pages} {074506} (\bibinfo {year} {2012})}\BibitemShut {NoStop}%
\bibitem [{\citenamefont {Shah}\ \emph {et~al.}(1999)\citenamefont {Shah}, \citenamefont {Kevrekidis},\ and\ \citenamefont {Benziger}}]{shah_ink-jet_1999}%
  \BibitemOpen
  \bibfield  {author} {\bibinfo {author} {\bibfnamefont {P.}~\bibnamefont {Shah}}, \bibinfo {author} {\bibfnamefont {Y.}~\bibnamefont {Kevrekidis}},\ and\ \bibinfo {author} {\bibfnamefont {J.}~\bibnamefont {Benziger}},\ }\bibfield  {title} {{\selectlanguage {en}\bibinfo {title} {Ink-jet printing of catalyst patterns for electroless metal deposition}},\ }\href {https://doi.org/10.1021/la9809123} {\bibfield  {journal} {\bibinfo  {journal} {Langmuir}\ }\textbf {\bibinfo {volume} {15}},\ \bibinfo {pages} {1584} (\bibinfo {year} {1999})}\BibitemShut {NoStop}%
\bibitem [{\citenamefont {Eggers}\ and\ \citenamefont {Villermaux}(2008)}]{eggers_physics_2008}%
  \BibitemOpen
  \bibfield  {author} {\bibinfo {author} {\bibfnamefont {J.}~\bibnamefont {Eggers}}\ and\ \bibinfo {author} {\bibfnamefont {E.}~\bibnamefont {Villermaux}},\ }\bibfield  {title} {{\selectlanguage {en}\bibinfo {title} {Physics of liquid jets}},\ }\href {https://doi.org/10.1088/0034-4885/71/3/036601} {\bibfield  {journal} {\bibinfo  {journal} {Rep. Prog. Phys.}\ }\textbf {\bibinfo {volume} {71}},\ \bibinfo {pages} {036601} (\bibinfo {year} {2008})}\BibitemShut {NoStop}%
\bibitem [{\citenamefont {Hoath}\ \emph {et~al.}(2013)\citenamefont {Hoath}, \citenamefont {Jung},\ and\ \citenamefont {Hutchings}}]{hoath_simple_2013}%
  \BibitemOpen
  \bibfield  {author} {\bibinfo {author} {\bibfnamefont {S.~D.}\ \bibnamefont {Hoath}}, \bibinfo {author} {\bibfnamefont {S.}~\bibnamefont {Jung}},\ and\ \bibinfo {author} {\bibfnamefont {I.~M.}\ \bibnamefont {Hutchings}},\ }\bibfield  {title} {{\selectlanguage {en}\bibinfo {title} {A simple criterion for filament break-up in drop-on-demand inkjet printing}},\ }\href {https://doi.org/10.1063/1.4790193} {\bibfield  {journal} {\bibinfo  {journal} {Phys. Fluids}\ }\textbf {\bibinfo {volume} {25}},\ \bibinfo {pages} {021701} (\bibinfo {year} {2013})}\BibitemShut {NoStop}%
\bibitem [{\citenamefont {Planchette}\ \emph {et~al.}(2019)\citenamefont {Planchette}, \citenamefont {Marangon}, \citenamefont {Hsiao},\ and\ \citenamefont {Brenn}}]{planchette_breakup_2019}%
  \BibitemOpen
  \bibfield  {author} {\bibinfo {author} {\bibfnamefont {C.}~\bibnamefont {Planchette}}, \bibinfo {author} {\bibfnamefont {F.}~\bibnamefont {Marangon}}, \bibinfo {author} {\bibfnamefont {W.-K.}\ \bibnamefont {Hsiao}},\ and\ \bibinfo {author} {\bibfnamefont {G.}~\bibnamefont {Brenn}},\ }\bibfield  {title} {{\selectlanguage {en}\bibinfo {title} {Breakup of asymmetric liquid ligaments}},\ }\href {https://doi.org/10.1103/PhysRevFluids.4.124004} {\bibfield  {journal} {\bibinfo  {journal} {Phys. Rev. Fluids}\ }\textbf {\bibinfo {volume} {4}},\ \bibinfo {pages} {124004} (\bibinfo {year} {2019})}\BibitemShut {NoStop}%
\bibitem [{\citenamefont {Schena}\ \emph {et~al.}(1998)\citenamefont {Schena}, \citenamefont {Heller}, \citenamefont {Theriault}, \citenamefont {Konrad}, \citenamefont {Lachenmeier},\ and\ \citenamefont {Davis}}]{schena_microarrays_1998}%
  \BibitemOpen
  \bibfield  {author} {\bibinfo {author} {\bibfnamefont {M.}~\bibnamefont {Schena}}, \bibinfo {author} {\bibfnamefont {R.~A.}\ \bibnamefont {Heller}}, \bibinfo {author} {\bibfnamefont {T.~P.}\ \bibnamefont {Theriault}}, \bibinfo {author} {\bibfnamefont {K.}~\bibnamefont {Konrad}}, \bibinfo {author} {\bibfnamefont {E.}~\bibnamefont {Lachenmeier}},\ and\ \bibinfo {author} {\bibfnamefont {R.~W.}\ \bibnamefont {Davis}},\ }\bibfield  {title} {{\selectlanguage {en}\bibinfo {title} {Microarrays: biotechnology's discovery platform for functional genomics}},\ }\href {https://doi.org/10.1016/S0167-7799(98)01219-0} {\bibfield  {journal} {\bibinfo  {journal} {Trends Biotechnol.}\ }\textbf {\bibinfo {volume} {16}},\ \bibinfo {pages} {301} (\bibinfo {year} {1998})}\BibitemShut {NoStop}%
\bibitem [{\citenamefont {Basaran}(2002)}]{basaran_small-scale_2002}%
  \BibitemOpen
  \bibfield  {author} {\bibinfo {author} {\bibfnamefont {O.~A.}\ \bibnamefont {Basaran}},\ }\bibfield  {title} {{\selectlanguage {en}\bibinfo {title} {Small-scale free surface flows with breakup: {Drop} formation and emerging applications}},\ }\href {https://doi.org/10.1002/aic.690480902} {\bibfield  {journal} {\bibinfo  {journal} {AIChE J.}\ }\textbf {\bibinfo {volume} {48}},\ \bibinfo {pages} {1842} (\bibinfo {year} {2002})}\BibitemShut {NoStop}%
\bibitem [{\citenamefont {Altieri}\ \emph {et~al.}(2014)\citenamefont {Altieri}, \citenamefont {Cryer},\ and\ \citenamefont {Acharya}}]{altieri_mechanisms_2014}%
  \BibitemOpen
  \bibfield  {author} {\bibinfo {author} {\bibfnamefont {A.}~\bibnamefont {Altieri}}, \bibinfo {author} {\bibfnamefont {S.~A.}\ \bibnamefont {Cryer}},\ and\ \bibinfo {author} {\bibfnamefont {L.}~\bibnamefont {Acharya}},\ }\bibfield  {title} {{\selectlanguage {en}\bibinfo {title} {Mechanisms, experiment, and theory of liquid sheet breakup and drop size from agricultural nozzles}},\ }\href {https://doi.org/10.1615/AtomizSpr.2014008779} {\bibfield  {journal} {\bibinfo  {journal} {Atomization Sprays}\ }\textbf {\bibinfo {volume} {24}},\ \bibinfo {pages} {695} (\bibinfo {year} {2014})}\BibitemShut {NoStop}%
\bibitem [{\citenamefont {Notz}\ and\ \citenamefont {Basaran}(2004)}]{notz_dynamics_2004}%
  \BibitemOpen
  \bibfield  {author} {\bibinfo {author} {\bibfnamefont {P.~K.}\ \bibnamefont {Notz}}\ and\ \bibinfo {author} {\bibfnamefont {O.~A.}\ \bibnamefont {Basaran}},\ }\bibfield  {title} {{\selectlanguage {en}\bibinfo {title} {Dynamics and breakup of a contracting liquid filament}},\ }\href {https://doi.org/10.1017/s0022112004009759} {\bibfield  {journal} {\bibinfo  {journal} {J. Fluid Mech.}\ }\textbf {\bibinfo {volume} {512}},\ \bibinfo {pages} {223} (\bibinfo {year} {2004})}\BibitemShut {NoStop}%
\bibitem [{\citenamefont {Ranz}(1959)}]{10.1063/1.1735095}%
  \BibitemOpen
  \bibfield  {author} {\bibinfo {author} {\bibfnamefont {W.~E.}\ \bibnamefont {Ranz}},\ }\bibfield  {title} {\bibinfo {title} {Some experiments on the dynamics of liquid films},\ }\href {https://doi.org/10.1063/1.1735095} {\bibfield  {journal} {\bibinfo  {journal} {J. Appl. Phys.}\ }\textbf {\bibinfo {volume} {30}},\ \bibinfo {pages} {1950} (\bibinfo {year} {1959})}\BibitemShut {NoStop}%
\bibitem [{\citenamefont {Hoepffner}\ and\ \citenamefont {Paré}(2013)}]{hoepffner_recoil_2013}%
  \BibitemOpen
  \bibfield  {author} {\bibinfo {author} {\bibfnamefont {J.}~\bibnamefont {Hoepffner}}\ and\ \bibinfo {author} {\bibfnamefont {G.}~\bibnamefont {Paré}},\ }\bibfield  {title} {{\selectlanguage {en}\bibinfo {title} {Recoil of a liquid filament: escape from pinch-off through creation of a vortex ring}},\ }\href {https://doi.org/10.1017/jfm.2013.472} {\bibfield  {journal} {\bibinfo  {journal} {J. Fluid Mech.}\ }\textbf {\bibinfo {volume} {734}},\ \bibinfo {pages} {183} (\bibinfo {year} {2013})}\BibitemShut {NoStop}%
\bibitem [{\citenamefont {Culick}(1960)}]{culick_1960}%
  \BibitemOpen
  \bibfield  {author} {\bibinfo {author} {\bibfnamefont {F.~E.~C.}\ \bibnamefont {Culick}},\ }\bibfield  {title} {\bibinfo {title} {Comments on a ruptured soap film},\ }\bibfield  {journal} {\bibinfo  {journal} {J. Appl. Phys.}\ }\textbf {\bibinfo {volume} {31}},\ \href {https://doi.org/10.1063/1.1735765} {10.1063/1.1735765} (\bibinfo {year} {1960})\BibitemShut {NoStop}%
\bibitem [{\citenamefont {Taylor}(1959)}]{taylor_1959}%
  \BibitemOpen
  \bibfield  {author} {\bibinfo {author} {\bibfnamefont {G.~I.}\ \bibnamefont {Taylor}},\ }\bibfield  {title} {\bibinfo {title} {The dynamics of thin sheets of fluid. {III}. disintegration of fluid sheets},\ }\href {https://doi.org/10.1098/rspa.1959.0196} {\bibfield  {journal} {\bibinfo  {journal} {Proc. R. Soc. A}\ }\textbf {\bibinfo {volume} {253}},\ \bibinfo {pages} {313} (\bibinfo {year} {1959})}\BibitemShut {NoStop}%
\bibitem [{\citenamefont {Savva}\ and\ \citenamefont {Bush}(2009)}]{Savva_Bush_2009}%
  \BibitemOpen
  \bibfield  {author} {\bibinfo {author} {\bibfnamefont {N.}~\bibnamefont {Savva}}\ and\ \bibinfo {author} {\bibfnamefont {J.~W.~M.}\ \bibnamefont {Bush}},\ }\bibfield  {title} {\bibinfo {title} {Viscous sheet retraction},\ }\href {https://doi.org/10.1017/S0022112009005795} {\bibfield  {journal} {\bibinfo  {journal} {J. Fluid Mech.}\ }\textbf {\bibinfo {volume} {626}},\ \bibinfo {pages} {211–240} (\bibinfo {year} {2009})}\BibitemShut {NoStop}%
\bibitem [{\citenamefont {Munro}\ and\ \citenamefont {Lister}(2018)}]{munro2018capillary}%
  \BibitemOpen
  \bibfield  {author} {\bibinfo {author} {\bibfnamefont {J.~P.}\ \bibnamefont {Munro}}\ and\ \bibinfo {author} {\bibfnamefont {J.~R.}\ \bibnamefont {Lister}},\ }\bibfield  {title} {\bibinfo {title} {Capillary retraction of the edge of a stretched viscous sheet},\ }\href@noop {} {\bibfield  {journal} {\bibinfo  {journal} {Journal of Fluid Mechanics}\ }\textbf {\bibinfo {volume} {844}},\ \bibinfo {pages} {R1} (\bibinfo {year} {2018})}\BibitemShut {NoStop}%
\bibitem [{\citenamefont {Sanjay}\ \emph {et~al.}(2022)\citenamefont {Sanjay}, \citenamefont {Sen}, \citenamefont {Kant},\ and\ \citenamefont {Lohse}}]{Sanjay_Sen_Kant_Lohse_2022}%
  \BibitemOpen
  \bibfield  {author} {\bibinfo {author} {\bibfnamefont {V.}~\bibnamefont {Sanjay}}, \bibinfo {author} {\bibfnamefont {U.}~\bibnamefont {Sen}}, \bibinfo {author} {\bibfnamefont {P.}~\bibnamefont {Kant}},\ and\ \bibinfo {author} {\bibfnamefont {D.}~\bibnamefont {Lohse}},\ }\bibfield  {title} {\bibinfo {title} {Taylor–culick retractions and the influence of the surroundings},\ }\href {https://doi.org/10.1017/jfm.2022.671} {\bibfield  {journal} {\bibinfo  {journal} {J. Fluid Mech.}\ }\textbf {\bibinfo {volume} {948}},\ \bibinfo {pages} {A14} (\bibinfo {year} {2022})}\BibitemShut {NoStop}%
\bibitem [{\citenamefont {Wee}\ \emph {et~al.}(2024)\citenamefont {Wee}, \citenamefont {Kumar}, \citenamefont {Liu},\ and\ \citenamefont {Basaran}}]{Wee_2024}%
  \BibitemOpen
  \bibfield  {author} {\bibinfo {author} {\bibfnamefont {H.}~\bibnamefont {Wee}}, \bibinfo {author} {\bibfnamefont {A.~H.}\ \bibnamefont {Kumar}}, \bibinfo {author} {\bibfnamefont {X.}~\bibnamefont {Liu}},\ and\ \bibinfo {author} {\bibfnamefont {O.~A.}\ \bibnamefont {Basaran}},\ }\bibfield  {title} {\bibinfo {title} {Escape from pinch-off during contraction of liquid sheets and two-dimensional drops of low-viscosity fluids},\ }\href {https://doi.org/10.1103/PhysRevFluids.9.103601} {\bibfield  {journal} {\bibinfo  {journal} {Phys. Rev. Fluids}\ }\textbf {\bibinfo {volume} {9}},\ \bibinfo {pages} {103601} (\bibinfo {year} {2024})}\BibitemShut {NoStop}%
\bibitem [{\citenamefont {Ahsan}\ \emph {et~al.}(2026)\citenamefont {Ahsan}, \citenamefont {Brand\~ao}, \citenamefont {Davidovitch},\ and\ \citenamefont {Stone}}]{Ahsan}%
  \BibitemOpen
  \bibfield  {author} {\bibinfo {author} {\bibfnamefont {T.}~\bibnamefont {Ahsan}}, \bibinfo {author} {\bibfnamefont {R.}~\bibnamefont {Brand\~ao}}, \bibinfo {author} {\bibfnamefont {B.}~\bibnamefont {Davidovitch}},\ and\ \bibinfo {author} {\bibfnamefont {H.~A.}\ \bibnamefont {Stone}},\ }\bibfield  {title} {\bibinfo {title} {Retraction dynamics of a highly viscous liquid sheet},\ }\href {https://doi.org/10.1103/fnhn-7zmw} {\bibfield  {journal} {\bibinfo  {journal} {Phys. Rev. Fluids}\ }\textbf {\bibinfo {volume} {11}},\ \bibinfo {pages} {014001} (\bibinfo {year} {2026})}\BibitemShut {NoStop}%
\bibitem [{\citenamefont {Constante-Amores}\ \emph {et~al.}(2020)\citenamefont {Constante-Amores}, \citenamefont {Kahouadji}, \citenamefont {Batchvarov}, \citenamefont {Shin}, \citenamefont {Chergui}, \citenamefont {Juric},\ and\ \citenamefont {Matar}}]{constante-amores_dynamics_2020}%
  \BibitemOpen
  \bibfield  {author} {\bibinfo {author} {\bibfnamefont {C.~R.}\ \bibnamefont {Constante-Amores}}, \bibinfo {author} {\bibfnamefont {L.}~\bibnamefont {Kahouadji}}, \bibinfo {author} {\bibfnamefont {A.}~\bibnamefont {Batchvarov}}, \bibinfo {author} {\bibfnamefont {S.}~\bibnamefont {Shin}}, \bibinfo {author} {\bibfnamefont {J.}~\bibnamefont {Chergui}}, \bibinfo {author} {\bibfnamefont {D.}~\bibnamefont {Juric}},\ and\ \bibinfo {author} {\bibfnamefont {O.~K.}\ \bibnamefont {Matar}},\ }\bibfield  {title} {{\selectlanguage {en}\bibinfo {title} {Dynamics of retracting surfactant-laden ligaments at intermediate {Ohnesorge} number}},\ }\href {https://doi.org/10.1103/PhysRevFluids.5.084007} {\bibfield  {journal} {\bibinfo  {journal} {Phys. Rev. Fluids}\ }\textbf {\bibinfo {volume} {5}},\ \bibinfo {pages} {084007} (\bibinfo {year} {2020})}\BibitemShut {NoStop}%
\bibitem [{\citenamefont {Kamat}\ \emph {et~al.}(2020)\citenamefont {Kamat}, \citenamefont {Wagoner}, \citenamefont {Castrejón-Pita}, \citenamefont {Castrejón-Pita}, \citenamefont {Anthony},\ and\ \citenamefont {Basaran}}]{kamat_surfactant-driven_2020}%
  \BibitemOpen
  \bibfield  {author} {\bibinfo {author} {\bibfnamefont {P.~M.}\ \bibnamefont {Kamat}}, \bibinfo {author} {\bibfnamefont {B.~W.}\ \bibnamefont {Wagoner}}, \bibinfo {author} {\bibfnamefont {A.~A.}\ \bibnamefont {Castrejón-Pita}}, \bibinfo {author} {\bibfnamefont {J.~R.}\ \bibnamefont {Castrejón-Pita}}, \bibinfo {author} {\bibfnamefont {C.~R.}\ \bibnamefont {Anthony}},\ and\ \bibinfo {author} {\bibfnamefont {O.~A.}\ \bibnamefont {Basaran}},\ }\bibfield  {title} {{\selectlanguage {en}\bibinfo {title} {Surfactant-driven escape from endpinching during contraction of nearly inviscid filaments}},\ }\href {https://doi.org/10.1017/jfm.2020.476} {\bibfield  {journal} {\bibinfo  {journal} {J. Fluid Mech.}\ }\textbf {\bibinfo {volume} {899}},\ \bibinfo {pages} {A28} (\bibinfo {year} {2020})}\BibitemShut {NoStop}%
\bibitem [{\citenamefont {De~Corato}\ \emph {et~al.}(2022)\citenamefont {De~Corato}, \citenamefont {Tammaro}, \citenamefont {Maffettone},\ and\ \citenamefont {Fueyo}}]{de_corato_retraction_2022}%
  \BibitemOpen
  \bibfield  {author} {\bibinfo {author} {\bibfnamefont {M.}~\bibnamefont {De~Corato}}, \bibinfo {author} {\bibfnamefont {D.}~\bibnamefont {Tammaro}}, \bibinfo {author} {\bibfnamefont {P.~L.}\ \bibnamefont {Maffettone}},\ and\ \bibinfo {author} {\bibfnamefont {N.}~\bibnamefont {Fueyo}},\ }\bibfield  {title} {{\selectlanguage {en}\bibinfo {title} {Retraction of thin films coated by insoluble surfactants}},\ }\href {https://doi.org/10.1017/jfm.2022.412} {\bibfield  {journal} {\bibinfo  {journal} {J. Fluid Mech.}\ }\textbf {\bibinfo {volume} {942}},\ \bibinfo {pages} {A55} (\bibinfo {year} {2022})}\BibitemShut {NoStop}%
\bibitem [{\citenamefont {Keshavarz}\ \emph {et~al.}(2015)\citenamefont {Keshavarz}, \citenamefont {Sharma}, \citenamefont {Houze}, \citenamefont {Koerner}, \citenamefont {Moore}, \citenamefont {Cotts}, \citenamefont {Threlfall-Holmes},\ and\ \citenamefont {McKinley}}]{keshavarz_studying_2015}%
  \BibitemOpen
  \bibfield  {author} {\bibinfo {author} {\bibfnamefont {B.}~\bibnamefont {Keshavarz}}, \bibinfo {author} {\bibfnamefont {V.}~\bibnamefont {Sharma}}, \bibinfo {author} {\bibfnamefont {E.~C.}\ \bibnamefont {Houze}}, \bibinfo {author} {\bibfnamefont {M.~R.}\ \bibnamefont {Koerner}}, \bibinfo {author} {\bibfnamefont {J.~R.}\ \bibnamefont {Moore}}, \bibinfo {author} {\bibfnamefont {P.~M.}\ \bibnamefont {Cotts}}, \bibinfo {author} {\bibfnamefont {P.}~\bibnamefont {Threlfall-Holmes}},\ and\ \bibinfo {author} {\bibfnamefont {G.~H.}\ \bibnamefont {McKinley}},\ }\bibfield  {title} {{\selectlanguage {en}\bibinfo {title} {Studying the effects of elongational properties on atomization of weakly viscoelastic solutions using {Rayleigh} {Ohnesorge} {Jetting} {Extensional} {Rheometry} ({ROJER})}},\ }\href {https://doi.org/10.1016/j.jnnfm.2014.11.004} {\bibfield  {journal} {\bibinfo  {journal} {J. Non-Newtonian Fluid Mech.}\ }\textbf {\bibinfo {volume} {222}},\ \bibinfo {pages} {171} (\bibinfo {year} {2015})}\BibitemShut {NoStop}%
\bibitem [{\citenamefont {Sen}\ \emph {et~al.}(2021)\citenamefont {Sen}, \citenamefont {Datt}, \citenamefont {Segers}, \citenamefont {Wijshoff}, \citenamefont {Snoeijer}, \citenamefont {Versluis},\ and\ \citenamefont {Lohse}}]{Sen_Datt_Segers_Wijshoff_Snoeijer_Versluis_Lohse_2021}%
  \BibitemOpen
  \bibfield  {author} {\bibinfo {author} {\bibfnamefont {U.}~\bibnamefont {Sen}}, \bibinfo {author} {\bibfnamefont {C.}~\bibnamefont {Datt}}, \bibinfo {author} {\bibfnamefont {T.}~\bibnamefont {Segers}}, \bibinfo {author} {\bibfnamefont {H.}~\bibnamefont {Wijshoff}}, \bibinfo {author} {\bibfnamefont {J.~H.}\ \bibnamefont {Snoeijer}}, \bibinfo {author} {\bibfnamefont {M.}~\bibnamefont {Versluis}},\ and\ \bibinfo {author} {\bibfnamefont {D.}~\bibnamefont {Lohse}},\ }\bibfield  {title} {\bibinfo {title} {The retraction of jetted slender viscoelastic liquid filaments},\ }\href {https://doi.org/10.1017/jfm.2021.855} {\bibfield  {journal} {\bibinfo  {journal} {J. Fluid Mech.}\ }\textbf {\bibinfo {volume} {929}},\ \bibinfo {pages} {A25} (\bibinfo {year} {2021})}\BibitemShut {NoStop}%
\bibitem [{\citenamefont {Liu}\ \emph {et~al.}(2022)\citenamefont {Liu}, \citenamefont {Wagoner},\ and\ \citenamefont {Basaran}}]{PhysRevFluids.7.L121601}%
  \BibitemOpen
  \bibfield  {author} {\bibinfo {author} {\bibfnamefont {X.}~\bibnamefont {Liu}}, \bibinfo {author} {\bibfnamefont {B.~W.}\ \bibnamefont {Wagoner}},\ and\ \bibinfo {author} {\bibfnamefont {O.~A.}\ \bibnamefont {Basaran}},\ }\bibfield  {title} {\bibinfo {title} {Contraction velocity of viscoelastic filaments},\ }\href {https://doi.org/10.1103/PhysRevFluids.7.L121601} {\bibfield  {journal} {\bibinfo  {journal} {Phys. Rev. Fluids}\ }\textbf {\bibinfo {volume} {7}},\ \bibinfo {pages} {L121601} (\bibinfo {year} {2022})}\BibitemShut {NoStop}%
\bibitem [{\citenamefont {Tammaro}\ \emph {et~al.}(2018{\natexlab{a}})\citenamefont {Tammaro}, \citenamefont {Pasquino}, \citenamefont {Villone}, \citenamefont {D’Avino}, \citenamefont {Ferraro}, \citenamefont {Di~Maio}, \citenamefont {Langella}, \citenamefont {Grizzuti},\ and\ \citenamefont {Maffettone}}]{tammaro2018elasticity}%
  \BibitemOpen
  \bibfield  {author} {\bibinfo {author} {\bibfnamefont {D.}~\bibnamefont {Tammaro}}, \bibinfo {author} {\bibfnamefont {R.}~\bibnamefont {Pasquino}}, \bibinfo {author} {\bibfnamefont {M.~M.}\ \bibnamefont {Villone}}, \bibinfo {author} {\bibfnamefont {G.}~\bibnamefont {D’Avino}}, \bibinfo {author} {\bibfnamefont {V.}~\bibnamefont {Ferraro}}, \bibinfo {author} {\bibfnamefont {E.}~\bibnamefont {Di~Maio}}, \bibinfo {author} {\bibfnamefont {A.}~\bibnamefont {Langella}}, \bibinfo {author} {\bibfnamefont {N.}~\bibnamefont {Grizzuti}},\ and\ \bibinfo {author} {\bibfnamefont {P.~L.}\ \bibnamefont {Maffettone}},\ }\bibfield  {title} {\bibinfo {title} {Elasticity in bubble rupture},\ }\href@noop {} {\bibfield  {journal} {\bibinfo  {journal} {Langmuir}\ }\textbf {\bibinfo {volume} {34}},\ \bibinfo {pages} {5646} (\bibinfo {year} {2018}{\natexlab{a}})}\BibitemShut {NoStop}%
\bibitem [{\citenamefont {Tammaro}\ \emph {et~al.}(2021{\natexlab{a}})\citenamefont {Tammaro}, \citenamefont {Chandran~Suja}, \citenamefont {Kannan}, \citenamefont {Gala}, \citenamefont {Di~Maio}, \citenamefont {Fuller},\ and\ \citenamefont {Maffettone}}]{tammaro2021flowering}%
  \BibitemOpen
  \bibfield  {author} {\bibinfo {author} {\bibfnamefont {D.}~\bibnamefont {Tammaro}}, \bibinfo {author} {\bibfnamefont {V.}~\bibnamefont {Chandran~Suja}}, \bibinfo {author} {\bibfnamefont {A.}~\bibnamefont {Kannan}}, \bibinfo {author} {\bibfnamefont {L.~D.}\ \bibnamefont {Gala}}, \bibinfo {author} {\bibfnamefont {E.}~\bibnamefont {Di~Maio}}, \bibinfo {author} {\bibfnamefont {G.~G.}\ \bibnamefont {Fuller}},\ and\ \bibinfo {author} {\bibfnamefont {P.~L.}\ \bibnamefont {Maffettone}},\ }\bibfield  {title} {\bibinfo {title} {Flowering in bursting bubbles with viscoelastic interfaces},\ }\href@noop {} {\bibfield  {journal} {\bibinfo  {journal} {Proceedings of the National Academy of Sciences}\ }\textbf {\bibinfo {volume} {118}},\ \bibinfo {pages} {e2105058118} (\bibinfo {year} {2021}{\natexlab{a}})}\BibitemShut {NoStop}%
\bibitem [{\citenamefont {Amoroso}\ and\ \citenamefont {Villone}(2025{\natexlab{a}})}]{amoroso2025numerical}%
  \BibitemOpen
  \bibfield  {author} {\bibinfo {author} {\bibfnamefont {D.}~\bibnamefont {Amoroso}}\ and\ \bibinfo {author} {\bibfnamefont {M.~M.}\ \bibnamefont {Villone}},\ }\bibfield  {title} {\bibinfo {title} {Numerical simulations of the stretching and relaxation dynamics of viscoelastic freestanding liquid films},\ }\href@noop {} {\bibfield  {journal} {\bibinfo  {journal} {Journal of Non-Newtonian Fluid Mechanics}\ ,\ \bibinfo {pages} {105544}} (\bibinfo {year} {2025}{\natexlab{a}})}\BibitemShut {NoStop}%
\bibitem [{\citenamefont {Deka}\ \emph {et~al.}(2019)\citenamefont {Deka}, \citenamefont {Pierson},\ and\ \citenamefont {Soares}}]{deka_retraction_2019}%
  \BibitemOpen
  \bibfield  {author} {\bibinfo {author} {\bibfnamefont {H.}~\bibnamefont {Deka}}, \bibinfo {author} {\bibfnamefont {J.-L.}\ \bibnamefont {Pierson}},\ and\ \bibinfo {author} {\bibfnamefont {E.~J.}\ \bibnamefont {Soares}},\ }\bibfield  {title} {{\selectlanguage {en}\bibinfo {title} {Retraction of a viscoplastic liquid sheet}},\ }\href {https://doi.org/10.1016/j.jnnfm.2019.104172} {\bibfield  {journal} {\bibinfo  {journal} {J. Non-Newtonian Fluid Mech.}\ }\textbf {\bibinfo {volume} {272}},\ \bibinfo {pages} {104172} (\bibinfo {year} {2019})}\BibitemShut {NoStop}%
\bibitem [{\citenamefont {Oppong}\ \emph {et~al.}(2006)\citenamefont {Oppong}, \citenamefont {Rubatat}, \citenamefont {Frisken}, \citenamefont {Bailey},\ and\ \citenamefont {de~Bruyn}}]{PhysRevE.73.041405}%
  \BibitemOpen
  \bibfield  {author} {\bibinfo {author} {\bibfnamefont {F.~K.}\ \bibnamefont {Oppong}}, \bibinfo {author} {\bibfnamefont {L.}~\bibnamefont {Rubatat}}, \bibinfo {author} {\bibfnamefont {B.~J.}\ \bibnamefont {Frisken}}, \bibinfo {author} {\bibfnamefont {A.~E.}\ \bibnamefont {Bailey}},\ and\ \bibinfo {author} {\bibfnamefont {J.~R.}\ \bibnamefont {de~Bruyn}},\ }\bibfield  {title} {\bibinfo {title} {Microrheology and structure of a yield-stress polymer gel},\ }\href {https://doi.org/10.1103/PhysRevE.73.041405} {\bibfield  {journal} {\bibinfo  {journal} {Phys. Rev. E}\ }\textbf {\bibinfo {volume} {73}},\ \bibinfo {pages} {041405} (\bibinfo {year} {2006})}\BibitemShut {NoStop}%
\bibitem [{\citenamefont {Saramito}(2009)}]{SARAMITO2009154}%
  \BibitemOpen
  \bibfield  {author} {\bibinfo {author} {\bibfnamefont {P.}~\bibnamefont {Saramito}},\ }\bibfield  {title} {\bibinfo {title} {A new elastoviscoplastic model based on the herschel–bulkley viscoplastic model},\ }\href {https://doi.org/https://doi.org/10.1016/j.jnnfm.2008.12.001} {\bibfield  {journal} {\bibinfo  {journal} {J. Non-Newtonian Fluid Mech.}\ }\textbf {\bibinfo {volume} {158}},\ \bibinfo {pages} {154} (\bibinfo {year} {2009})},\ \bibinfo {note} {visco-plastic fluids: From theory to application}\BibitemShut {NoStop}%
\bibitem [{\citenamefont {França}\ \emph {et~al.}(2026)\citenamefont {França}, \citenamefont {Tieman}, \citenamefont {Shemilt}, \citenamefont {Oishi},\ and\ \citenamefont {Jalaal}}]{franca2026coalescenceprintedyieldstress}%
  \BibitemOpen
  \bibfield  {author} {\bibinfo {author} {\bibfnamefont {H.~L.}\ \bibnamefont {França}}, \bibinfo {author} {\bibfnamefont {D.}~\bibnamefont {Tieman}}, \bibinfo {author} {\bibfnamefont {J.~D.}\ \bibnamefont {Shemilt}}, \bibinfo {author} {\bibfnamefont {C.}~\bibnamefont {Oishi}},\ and\ \bibinfo {author} {\bibfnamefont {M.}~\bibnamefont {Jalaal}},\ }\href {https://arxiv.org/abs/2601.09870} {\bibinfo {title} {Coalescence of printed yield stress filaments in direct ink writing}} (\bibinfo {year} {2026}),\ \Eprint {https://arxiv.org/abs/2601.09870} {arXiv:2601.09870 [cond-mat.soft]} \BibitemShut {NoStop}%
\bibitem [{\citenamefont {Zakeri}\ \emph {et~al.}(2025)\citenamefont {Zakeri}, \citenamefont {Moschopoulos}, \citenamefont {Dimakopoulos},\ and\ \citenamefont {Tsamopoulos}}]{Zakeri_Moschopoulos_Dimakopoulos_Tsamopoulos_2025}%
  \BibitemOpen
  \bibfield  {author} {\bibinfo {author} {\bibfnamefont {P.}~\bibnamefont {Zakeri}}, \bibinfo {author} {\bibfnamefont {P.}~\bibnamefont {Moschopoulos}}, \bibinfo {author} {\bibfnamefont {Y.}~\bibnamefont {Dimakopoulos}},\ and\ \bibinfo {author} {\bibfnamefont {J.}~\bibnamefont {Tsamopoulos}},\ }\bibfield  {title} {\bibinfo {title} {Scaling analysis and self-similarity near breakup of elasto-viscoplastic liquid threads under creeping flow},\ }\href {https://doi.org/10.1017/jfm.2025.10652} {\bibfield  {journal} {\bibinfo  {journal} {Journal of Fluid Mechanics}\ }\textbf {\bibinfo {volume} {1020}},\ \bibinfo {pages} {A37} (\bibinfo {year} {2025})}\BibitemShut {NoStop}%
\bibitem [{\citenamefont {Tammaro}\ \emph {et~al.}(2018{\natexlab{b}})\citenamefont {Tammaro}, \citenamefont {Pasquino}, \citenamefont {Villone}, \citenamefont {D’Avino}, \citenamefont {Ferraro}, \citenamefont {Di~Maio}, \citenamefont {Langella}, \citenamefont {Grizzuti},\ and\ \citenamefont {Maffettone}}]{doi:10.1021/acs.langmuir.8b00520}%
  \BibitemOpen
  \bibfield  {author} {\bibinfo {author} {\bibfnamefont {D.}~\bibnamefont {Tammaro}}, \bibinfo {author} {\bibfnamefont {R.}~\bibnamefont {Pasquino}}, \bibinfo {author} {\bibfnamefont {M.~M.}\ \bibnamefont {Villone}}, \bibinfo {author} {\bibfnamefont {G.}~\bibnamefont {D’Avino}}, \bibinfo {author} {\bibfnamefont {V.}~\bibnamefont {Ferraro}}, \bibinfo {author} {\bibfnamefont {E.}~\bibnamefont {Di~Maio}}, \bibinfo {author} {\bibfnamefont {A.}~\bibnamefont {Langella}}, \bibinfo {author} {\bibfnamefont {N.}~\bibnamefont {Grizzuti}},\ and\ \bibinfo {author} {\bibfnamefont {P.~L.}\ \bibnamefont {Maffettone}},\ }\bibfield  {title} {\bibinfo {title} {Elasticity in bubble rupture},\ }\href {https://doi.org/10.1021/acs.langmuir.8b00520} {\bibfield  {journal} {\bibinfo  {journal} {Langmuir}\ }\textbf {\bibinfo {volume} {34}},\ \bibinfo {pages} {5646} (\bibinfo {year} {2018}{\natexlab{b}})},\ \bibinfo {note} {pMID: 29664652}\BibitemShut {NoStop}%
\bibitem [{\citenamefont {Tammaro}\ \emph {et~al.}(2021{\natexlab{b}})\citenamefont {Tammaro}, \citenamefont {Suja}, \citenamefont {Kannan}, \citenamefont {Gala}, \citenamefont {Maio}, \citenamefont {Fuller},\ and\ \citenamefont {Maffettone}}]{doi:10.1073/pnas.2105058118}%
  \BibitemOpen
  \bibfield  {author} {\bibinfo {author} {\bibfnamefont {D.}~\bibnamefont {Tammaro}}, \bibinfo {author} {\bibfnamefont {V.~C.}\ \bibnamefont {Suja}}, \bibinfo {author} {\bibfnamefont {A.}~\bibnamefont {Kannan}}, \bibinfo {author} {\bibfnamefont {L.~D.}\ \bibnamefont {Gala}}, \bibinfo {author} {\bibfnamefont {E.~D.}\ \bibnamefont {Maio}}, \bibinfo {author} {\bibfnamefont {G.~G.}\ \bibnamefont {Fuller}},\ and\ \bibinfo {author} {\bibfnamefont {P.~L.}\ \bibnamefont {Maffettone}},\ }\bibfield  {title} {\bibinfo {title} {Flowering in bursting bubbles with viscoelastic interfaces},\ }\href {https://doi.org/10.1073/pnas.2105058118} {\bibfield  {journal} {\bibinfo  {journal} {Proceedings of the National Academy of Sciences}\ }\textbf {\bibinfo {volume} {118}},\ \bibinfo {pages} {e2105058118} (\bibinfo {year} {2021}{\natexlab{b}})},\ \Eprint {https://arxiv.org/abs/https://www.pnas.org/doi/pdf/10.1073/pnas.2105058118} {https://www.pnas.org/doi/pdf/10.1073/pnas.2105058118} \BibitemShut {NoStop}%
\bibitem [{\citenamefont {Amoroso}\ and\ \citenamefont {Villone}(2025{\natexlab{b}})}]{Amoroso2025NumericalSO}%
  \BibitemOpen
  \bibfield  {author} {\bibinfo {author} {\bibfnamefont {D.}~\bibnamefont {Amoroso}}\ and\ \bibinfo {author} {\bibfnamefont {M.~M.}\ \bibnamefont {Villone}},\ }\bibfield  {title} {\bibinfo {title} {Numerical simulations of the stretching and relaxation dynamics of viscoelastic freestanding liquid films},\ }\href {https://api.semanticscholar.org/CorpusID:284114097} {\bibfield  {journal} {\bibinfo  {journal} {Journal of Non-Newtonian Fluid Mechanics}\ } (\bibinfo {year} {2025}{\natexlab{b}})}\BibitemShut {NoStop}%
\bibitem [{\citenamefont {Yao}\ \emph {et~al.}(2025)\citenamefont {Yao}, \citenamefont {Gao}, \citenamefont {Yu}, \citenamefont {Qi}, \citenamefont {Lin}, \citenamefont {Xu}, \citenamefont {Li}, \citenamefont {Zhang}, \citenamefont {Zhu},\ and\ \citenamefont {Lu}}]{YAO2025113550}%
  \BibitemOpen
  \bibfield  {author} {\bibinfo {author} {\bibfnamefont {Y.}~\bibnamefont {Yao}}, \bibinfo {author} {\bibfnamefont {Q.}~\bibnamefont {Gao}}, \bibinfo {author} {\bibfnamefont {K.}~\bibnamefont {Yu}}, \bibinfo {author} {\bibfnamefont {L.}~\bibnamefont {Qi}}, \bibinfo {author} {\bibfnamefont {Z.}~\bibnamefont {Lin}}, \bibinfo {author} {\bibfnamefont {J.}~\bibnamefont {Xu}}, \bibinfo {author} {\bibfnamefont {Y.}~\bibnamefont {Li}}, \bibinfo {author} {\bibfnamefont {P.}~\bibnamefont {Zhang}}, \bibinfo {author} {\bibfnamefont {M.}~\bibnamefont {Zhu}},\ and\ \bibinfo {author} {\bibfnamefont {L.}~\bibnamefont {Lu}},\ }\bibfield  {title} {\bibinfo {title} {Study on the formation mechanism of viscoplastic line deposition for predicting filament width},\ }\href@noop {} {\bibfield  {journal} {\bibinfo  {journal} {Materials \& Design}\ }\textbf {\bibinfo {volume} {249}},\ \bibinfo {pages} {113550} (\bibinfo {year} {2025})}\BibitemShut {NoStop}%
\bibitem [{\citenamefont {Ball}\ and\ \citenamefont {Balmforth}(2024)}]{HB_reference}%
  \BibitemOpen
  \bibfield  {author} {\bibinfo {author} {\bibfnamefont {T.~V.}\ \bibnamefont {Ball}}\ and\ \bibinfo {author} {\bibfnamefont {N.~J.}\ \bibnamefont {Balmforth}},\ }\bibfield  {title} {\bibinfo {title} {Viscoplastic rimming flow inside a rotating cylinder},\ }\href {https://doi.org/10.1103/PhysRevFluids.9.023304} {\bibfield  {journal} {\bibinfo  {journal} {Phys. Rev. Fluids}\ }\textbf {\bibinfo {volume} {9}},\ \bibinfo {pages} {023304} (\bibinfo {year} {2024})}\BibitemShut {NoStop}%
\bibitem [{\citenamefont {Matoba}\ \emph {et~al.}(2025)\citenamefont {Matoba}, \citenamefont {Iceri}, \citenamefont {{de Moura}}, \citenamefont {Thompson}, \citenamefont {Fidel-Dufour}, \citenamefont {Palermo},\ and\ \citenamefont {{de Castro}}}]{MATOBA2025105481}%
  \BibitemOpen
  \bibfield  {author} {\bibinfo {author} {\bibfnamefont {G.~K.}\ \bibnamefont {Matoba}}, \bibinfo {author} {\bibfnamefont {D.~M.}\ \bibnamefont {Iceri}}, \bibinfo {author} {\bibfnamefont {H.~L.}\ \bibnamefont {{de Moura}}}, \bibinfo {author} {\bibfnamefont {R.~L.}\ \bibnamefont {Thompson}}, \bibinfo {author} {\bibfnamefont {A.}~\bibnamefont {Fidel-Dufour}}, \bibinfo {author} {\bibfnamefont {T.}~\bibnamefont {Palermo}},\ and\ \bibinfo {author} {\bibfnamefont {M.~S.}\ \bibnamefont {{de Castro}}},\ }\bibfield  {title} {\bibinfo {title} {Evaluating viscoplastic properties with rheometry and piv measurements in pipeline flows},\ }\href@noop {} {\bibfield  {journal} {\bibinfo  {journal} {Journal of Non-Newtonian Fluid Mechanics}\ }\textbf {\bibinfo {volume} {345}},\ \bibinfo {pages} {105481} (\bibinfo {year} {2025})}\BibitemShut {NoStop}%
\bibitem [{\citenamefont {Papanastasiou}(1987)}]{10.1122/1.549926}%
  \BibitemOpen
  \bibfield  {author} {\bibinfo {author} {\bibfnamefont {T.~C.}\ \bibnamefont {Papanastasiou}},\ }\bibfield  {title} {\bibinfo {title} {Flows of materials with yield},\ }\href {https://doi.org/10.1122/1.549926} {\bibfield  {journal} {\bibinfo  {journal} {J. Rheol.}\ }\textbf {\bibinfo {volume} {31}},\ \bibinfo {pages} {385} (\bibinfo {year} {1987})}\BibitemShut {NoStop}%
\bibitem [{\citenamefont {Popinet}(2009{\natexlab{a}})}]{popinet_accurate_2009}%
  \BibitemOpen
  \bibfield  {author} {\bibinfo {author} {\bibfnamefont {S.}~\bibnamefont {Popinet}},\ }\bibfield  {title} {\bibinfo {title} {An accurate adaptive solver for surface-tension-driven interfacial flows},\ }\href {https://doi.org/10.1016/j.jcp.2009.04.042} {\bibfield  {journal} {\bibinfo  {journal} {J. Comput. Phys.}\ }\textbf {\bibinfo {volume} {228}},\ \bibinfo {pages} {5838} (\bibinfo {year} {2009}{\natexlab{a}})}\BibitemShut {NoStop}%
\bibitem [{\citenamefont {Popinet}\ \emph {et~al.}(2025)\citenamefont {Popinet} \emph {et~al.}}]{popinet2013basilisk}%
  \BibitemOpen
  \bibfield  {author} {\bibinfo {author} {\bibfnamefont {S.}~\bibnamefont {Popinet}} \emph {et~al.},\ }\href@noop {} {\bibinfo {title} {Basilisk}},\ \bibinfo {howpublished} {\url{http://basilisk.fr}} (\bibinfo {year} {2013--2025})\BibitemShut {NoStop}%
\bibitem [{\citenamefont {Sanjay}\ \emph {et~al.}(2021)\citenamefont {Sanjay}, \citenamefont {Lohse},\ and\ \citenamefont {Jalaal}}]{sanjay_bursting_2021}%
  \BibitemOpen
  \bibfield  {author} {\bibinfo {author} {\bibfnamefont {V.}~\bibnamefont {Sanjay}}, \bibinfo {author} {\bibfnamefont {D.}~\bibnamefont {Lohse}},\ and\ \bibinfo {author} {\bibfnamefont {M.}~\bibnamefont {Jalaal}},\ }\bibfield  {title} {{\selectlanguage {en}\bibinfo {title} {Bursting bubble in a viscoplastic medium}},\ }\href {https://doi.org/10.1017/jfm.2021.489} {\bibfield  {journal} {\bibinfo  {journal} {J. Fluid Mech.}\ }\textbf {\bibinfo {volume} {922}},\ \bibinfo {pages} {A2} (\bibinfo {year} {2021})}\BibitemShut {NoStop}%
\bibitem [{\citenamefont {Van~Hooft}\ \emph {et~al.}(2018)\citenamefont {Van~Hooft}, \citenamefont {Popinet}, \citenamefont {Van~Heerwaarden}, \citenamefont {Van Der~Linden}, \citenamefont {De~Roode},\ and\ \citenamefont {Van De~Wiel}}]{van_hooft_towards_2018}%
  \BibitemOpen
  \bibfield  {author} {\bibinfo {author} {\bibfnamefont {J.~A.}\ \bibnamefont {Van~Hooft}}, \bibinfo {author} {\bibfnamefont {S.}~\bibnamefont {Popinet}}, \bibinfo {author} {\bibfnamefont {C.~C.}\ \bibnamefont {Van~Heerwaarden}}, \bibinfo {author} {\bibfnamefont {S.~J.~A.}\ \bibnamefont {Van Der~Linden}}, \bibinfo {author} {\bibfnamefont {S.~R.}\ \bibnamefont {De~Roode}},\ and\ \bibinfo {author} {\bibfnamefont {B.~J.~H.}\ \bibnamefont {Van De~Wiel}},\ }\bibfield  {title} {{\selectlanguage {en}\bibinfo {title} {Towards adaptive grids for atmospheric boundary-layer simulations}},\ }\href {https://doi.org/10.1007/s10546-018-0335-9} {\bibfield  {journal} {\bibinfo  {journal} {Boundary-Layer Meteorol.}\ }\textbf {\bibinfo {volume} {167}},\ \bibinfo {pages} {421} (\bibinfo {year} {2018})}\BibitemShut {NoStop}%
\bibitem [{\citenamefont {Day}\ \emph {et~al.}(1998)\citenamefont {Day}, \citenamefont {Hinch},\ and\ \citenamefont {Lister}}]{day_self-similar_1998}%
  \BibitemOpen
  \bibfield  {author} {\bibinfo {author} {\bibfnamefont {R.~F.}\ \bibnamefont {Day}}, \bibinfo {author} {\bibfnamefont {E.~J.}\ \bibnamefont {Hinch}},\ and\ \bibinfo {author} {\bibfnamefont {J.~R.}\ \bibnamefont {Lister}},\ }\bibfield  {title} {{\selectlanguage {en}\bibinfo {title} {Self-similar capillary pinchoff of an inviscid fluid}},\ }\href {https://doi.org/10.1103/PhysRevLett.80.704} {\bibfield  {journal} {\bibinfo  {journal} {Phys. Rev. Lett.}\ }\textbf {\bibinfo {volume} {80}},\ \bibinfo {pages} {704} (\bibinfo {year} {1998})}\BibitemShut {NoStop}%
\bibitem [{\citenamefont {Pierson}\ \emph {et~al.}(2020)\citenamefont {Pierson}, \citenamefont {Magnaudet}, \citenamefont {Soares},\ and\ \citenamefont {Popinet}}]{PhysRevFluids.5.073602}%
  \BibitemOpen
  \bibfield  {author} {\bibinfo {author} {\bibfnamefont {J.-L.}\ \bibnamefont {Pierson}}, \bibinfo {author} {\bibfnamefont {J.}~\bibnamefont {Magnaudet}}, \bibinfo {author} {\bibfnamefont {E.~J.}\ \bibnamefont {Soares}},\ and\ \bibinfo {author} {\bibfnamefont {S.}~\bibnamefont {Popinet}},\ }\bibfield  {title} {\bibinfo {title} {Revisiting the taylor-culick approximation: Retraction of an axisymmetric filament},\ }\href {https://doi.org/10.1103/PhysRevFluids.5.073602} {\bibfield  {journal} {\bibinfo  {journal} {Phys. Rev. Fluids}\ }\textbf {\bibinfo {volume} {5}},\ \bibinfo {pages} {073602} (\bibinfo {year} {2020})}\BibitemShut {NoStop}%
\bibitem [{\citenamefont {Brenner}\ and\ \citenamefont {Gueyffier}(1999)}]{10.1063/1.869942}%
  \BibitemOpen
  \bibfield  {author} {\bibinfo {author} {\bibfnamefont {M.~P.}\ \bibnamefont {Brenner}}\ and\ \bibinfo {author} {\bibfnamefont {D.}~\bibnamefont {Gueyffier}},\ }\bibfield  {title} {\bibinfo {title} {On the bursting of viscous films},\ }\href {https://doi.org/10.1063/1.869942} {\bibfield  {journal} {\bibinfo  {journal} {Phys. Fluids}\ }\textbf {\bibinfo {volume} {11}},\ \bibinfo {pages} {737} (\bibinfo {year} {1999})}\BibitemShut {NoStop}%
\bibitem [{\citenamefont {Castrejón-Pita}\ \emph {et~al.}(2012)\citenamefont {Castrejón-Pita}, \citenamefont {Castrejón-Pita},\ and\ \citenamefont {Hutchings}}]{castrejon-pita_breakup_2012}%
  \BibitemOpen
  \bibfield  {author} {\bibinfo {author} {\bibfnamefont {A.~A.}\ \bibnamefont {Castrejón-Pita}}, \bibinfo {author} {\bibfnamefont {J.~R.}\ \bibnamefont {Castrejón-Pita}},\ and\ \bibinfo {author} {\bibfnamefont {I.~M.}\ \bibnamefont {Hutchings}},\ }\bibfield  {title} {{\selectlanguage {en}\bibinfo {title} {Breakup of liquid filaments}},\ }\href {https://doi.org/10.1103/PhysRevLett.108.074506} {\bibfield  {journal} {\bibinfo  {journal} {Phys. Rev. Lett.}\ }\textbf {\bibinfo {volume} {108}},\ \bibinfo {pages} {074506} (\bibinfo {year} {2012})}\BibitemShut {NoStop}%
\bibitem [{\citenamefont {Hossain}\ \emph {et~al.}(2025)\citenamefont {Hossain}, \citenamefont {Eom}, \citenamefont {Shah}, \citenamefont {Lowe}, \citenamefont {Fudge}, \citenamefont {Tawfick},\ and\ \citenamefont {Ewoldt}}]{tanver}%
  \BibitemOpen
  \bibfield  {author} {\bibinfo {author} {\bibfnamefont {M.~T.}\ \bibnamefont {Hossain}}, \bibinfo {author} {\bibfnamefont {W.}~\bibnamefont {Eom}}, \bibinfo {author} {\bibfnamefont {A.}~\bibnamefont {Shah}}, \bibinfo {author} {\bibfnamefont {A.}~\bibnamefont {Lowe}}, \bibinfo {author} {\bibfnamefont {D.}~\bibnamefont {Fudge}}, \bibinfo {author} {\bibfnamefont {S.~H.}\ \bibnamefont {Tawfick}},\ and\ \bibinfo {author} {\bibfnamefont {R.~H.}\ \bibnamefont {Ewoldt}},\ }\bibfield  {title} {\bibinfo {title} {The critical plastocapillary number for a {N}ewtonian liquid filament embedded into a viscoplastic fluid},\ }\href@noop {} {\bibfield  {journal} {\bibinfo  {journal} {J. Non-Newtonian Fluid Mech.}\ ,\ \bibinfo {pages} {105440}} (\bibinfo {year} {2025})}\BibitemShut {NoStop}%
\bibitem [{\citenamefont {Popinet}(2009{\natexlab{b}})}]{popinet2009accurate}%
  \BibitemOpen
  \bibfield  {author} {\bibinfo {author} {\bibfnamefont {S.}~\bibnamefont {Popinet}},\ }\bibfield  {title} {\bibinfo {title} {An accurate adaptive solver for surface-tension-driven interfacial flows},\ }\href@noop {} {\bibfield  {journal} {\bibinfo  {journal} {Journal of Computational Physics}\ }\textbf {\bibinfo {volume} {228}},\ \bibinfo {pages} {5838} (\bibinfo {year} {2009}{\natexlab{b}})}\BibitemShut {NoStop}%
\bibitem [{\citenamefont {Popinet}(nd)}]{BasiliskPlateauTest}%
  \BibitemOpen
  \bibfield  {author} {\bibinfo {author} {\bibfnamefont {S.}~\bibnamefont {Popinet}},\ }\href@noop {} {\bibinfo {title} {Rayleigh--plateau instability test case (plateau.c)}},\ \bibinfo {howpublished} {\url{https://basilisk.fr/src/test/plateau.c}} (\bibinfo {year} {n.d.}),\ \bibinfo {note} {accessed: 2026-03-31}\BibitemShut {NoStop}%
\bibitem [{\citenamefont {Panaseti}\ \emph {et~al.}(2018)\citenamefont {Panaseti}, \citenamefont {Damianou}, \citenamefont {Georgiou},\ and\ \citenamefont {Housiadas}}]{panaseti2018pressure}%
  \BibitemOpen
  \bibfield  {author} {\bibinfo {author} {\bibfnamefont {P.}~\bibnamefont {Panaseti}}, \bibinfo {author} {\bibfnamefont {Y.}~\bibnamefont {Damianou}}, \bibinfo {author} {\bibfnamefont {G.~C.}\ \bibnamefont {Georgiou}},\ and\ \bibinfo {author} {\bibfnamefont {K.~D.}\ \bibnamefont {Housiadas}},\ }\bibfield  {title} {\bibinfo {title} {Pressure-driven flow of a herschel-bulkley fluid with pressure-dependent rheological parameters},\ }\href@noop {} {\bibfield  {journal} {\bibinfo  {journal} {Phys. Fluids}\ }\textbf {\bibinfo {volume} {30}} (\bibinfo {year} {2018})}\BibitemShut {NoStop}%
\end{thebibliography}%

\end{document}